\documentclass[12pt,preprint,linenumbers]{article}

\usepackage{times}
\usepackage{lineno}

\usepackage[superscript]{cite}

\usepackage{graphicx}
\usepackage{epstopdf}
\usepackage{amssymb}
\DeclareGraphicsExtensions{.eps,.jpg,.png,.pdf}
\usepackage{caption}
\usepackage{setspace}
\title{On the Structure of Liquids: More order than expected}

\author
{Zhen Zhang,$^{1}$ Walter Kob,$^{1\ast}$ \\
\\
\normalsize{$^{1}$Laboratoire Charles Coulomb, University of Montpellier and CNRS,}\\
\normalsize{F-34095 Montpellier, France}\\
\\
\normalsize{$^\ast$Correspondence to:  walter.kob@umontpellier.fr}
}

\date{}

\begin{document} 




\maketitle

\begin{abstract}
Disordered systems like liquids, gels, glasses, or granular materials
are not only ubiquitous in daily life and in industrial applications
but they are also crucial for the mechanical stability of cells or
the transport of chemical and biological agents in living organisms
\cite{binder_11,cosgrove_05,becker_13}. Despite the importance of
these systems, their microscopic structure is understood only on a
rudimentary level, thus in stark contrast to the case of gases and
crystals~\cite{binder_11,ashcroft_76}. Experimental and theoretical
investigations indicate that disordered systems have a structural order on the
length scale of a few particle diameters but which then quickly vanishes
at larger distances~\cite{hansen_86}. This conclusion is, however, based
mainly on the behavior of two-point correlation functions such as the static
structure factor or the radial distribution function~\cite{hansen_86,salmon_06}.
Whether or not disordered systems have an order that extents to larger
length scales is therefore a highly important question that is still
open~\cite{tanaka_12,royall_15,schoenholz_16,royall_17}.  Here we use
computer simulations to show that liquids have an intricate structural
order given by alternating layers with icosahedral and dodecahedral
symmetries and which extends to surprisingly large distances. We show
that the temperature dependence of the corresponding length scale can be detected
in the static structure factor, making it directly accessible to
scattering experiments. These results, obtained by a novel type of four
point correlation function that probes directly the particle density in
three dimensions, show that liquids are far more ordered than envisaged
so far but that this structural information is encoded in non-standard
correlation functions.

\end{abstract}

The structure of complex systems is usually determined from scattering
experiments which give access to the static structure factor
$S(\vec{q})$, $\vec{q}$ is the wave-vector, and for crystals such
measurements allow to obtain a complete know\-ledge of the structure of
the material~\cite{hansen_86,salmon_06,ashcroft_76}. This is not the case
for disordered materials since these are isotropic and hence $S(\vec{q})$
depends only on the norm $q=|\vec{q}|$, i.e., the whole three dimensional
structural information is projected onto a single function $S(q)$. This
projection entails a huge loss of structural information,
which subsequently has to be recovered, at least partially,
from physical arguments on the possible arrangement of the
particles. Since for intermediate and large length scales no
such arguments exist, our current knowledge about the packing of
the particles is restricted to length scales of 2-3 particle diameters
\cite{jonsson_88,ma_11,miracle_04,wochner_09,xia_17,royall_15,malins_13,dunleavy_15,coslovich_07,tanaka_12}
whereas there is very little insight regarding the structure on larger
scales.

Whether or not disordered systems have indeed a structural
order that extents beyond a few particle diameters is one
of the central questions of glass science since various
theoretical approaches connect the slowing down of the
relaxation dynamics to the presence of an increasing static length
scale~\cite{binder_11,royall_17,rfot,adam_65}, although this mechanism
is challenged by other theories~\cite{chandler_10}. As a consequence
there have been a multitude of proposals regarding the nature and
definition of such a growing length, but so far no consensus has emerged
regarding the best choice, notably in the generic case of multi-component
systems~\cite{royall_17,dunleavy_15,coslovich_07,tanaka_12,fang_10,fang_11}.
In the present work we use a novel approach to reveal that liquids do
have highly non-trivial correlations up to distances well beyond the
first few neighbors.

\begin{figure}[ht]
\center
\includegraphics[width=0.3\columnwidth]{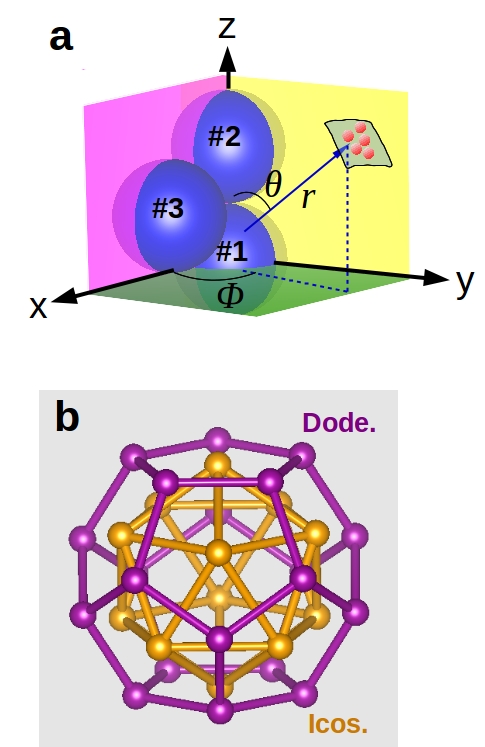}
\includegraphics[width=0.5\columnwidth]{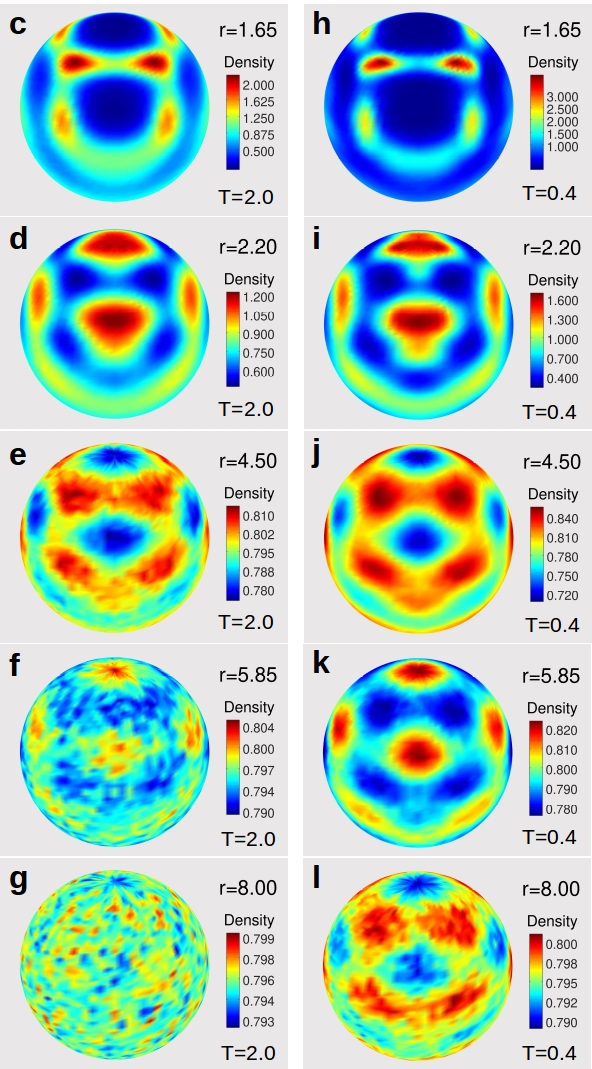}
\caption{
{\bf Distribution of particles in three dimensions.} 
{\bf a}: The definition of the local coordinate system
involves three A particles that are nearest neighbors to each other. {\bf b}:
An icosahedron is the dual polyhedron of a dodecahedron and vice
versa. {\bf c} to {\bf l}: Density distribution $\rho(\theta,\phi,r$) for
different values of $r$, i.e., the distribution of the particles that are
in a spherical shell of radius $r$ and thickness 0.4 around the central
particle. $T=2.0$ ({\bf c} to {\bf g}) and $T=0.4$ ({\bf h} to {\bf l}). 
In the reddish areas the density is high and in the bluish
areas the density is low.  Depending on the distance $r$ the high density regions
show an icosahedral ({\bf d}, {\bf i}, {\bf f}, and {\bf k}) or dodecahedral symmetry
({\bf c}, {\bf h}, {\bf e}, {\bf j}, and {\bf l}). 
}
\label{fig1_densityplot}
\end{figure}

The system we consider is a binary mixture of Lennard-Jones
particles (80\% A particles and 20\% B particles) that is not prone to
crystallization even at low temperatures (see Methods)~\cite{kob_95}. We
study the equilibrium properties of this liquid in a temperature range in
which the system changes from a very fluid state to a moderately viscous
one, i.e.~$5.0\geq T \geq 0.40$. To probe the three dimensional structure
of the system we introduce a local coordinate system as follows: Take any
three A particles that touch each other, i.e., they form a triangle with
sides that are less than the location of the first minimum in the radial
distribution function $g(r)$, i.e.$\approx 1.4$ (see~\ref{SI_fig_gofr}). We
define the position of particle \#1 as the origin, the direction from
particle \#1 to \#2 as the $z-$axis, and the plane containing the three
particles as the $z-x-$plane (Fig.~\ref{fig1_densityplot}{\bf a}). This
local reference frame allows to introduce a spherical coordinate
system $\theta,\phi,r$ and to measure the probability of finding any
other particle at a given point in space, i.e.~to measure a four point
correlation function.  Note that this coordinate system can be defined for
all triplets of neighboring particles and these density distributions can
be averaged to improve the statistics. Since this coordinate system is
adapted to the configuration by the three particles it allows to detect
angular correlations that are not visible in $g(r)$ or in previously
considered structural observables.

\begin{figure}[ht]
\center
\includegraphics[width=0.48\columnwidth]{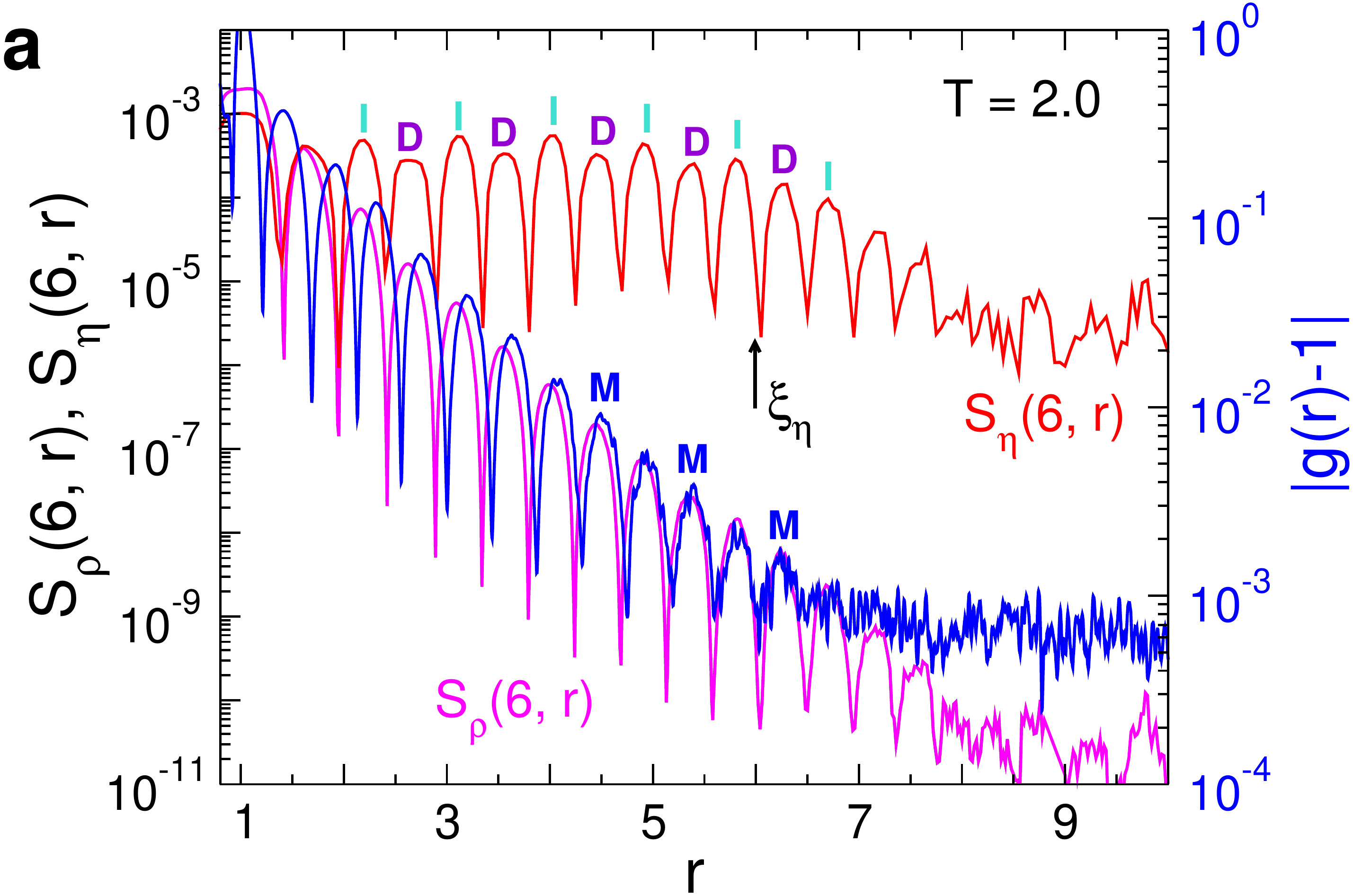}
\includegraphics[width=0.48\columnwidth]{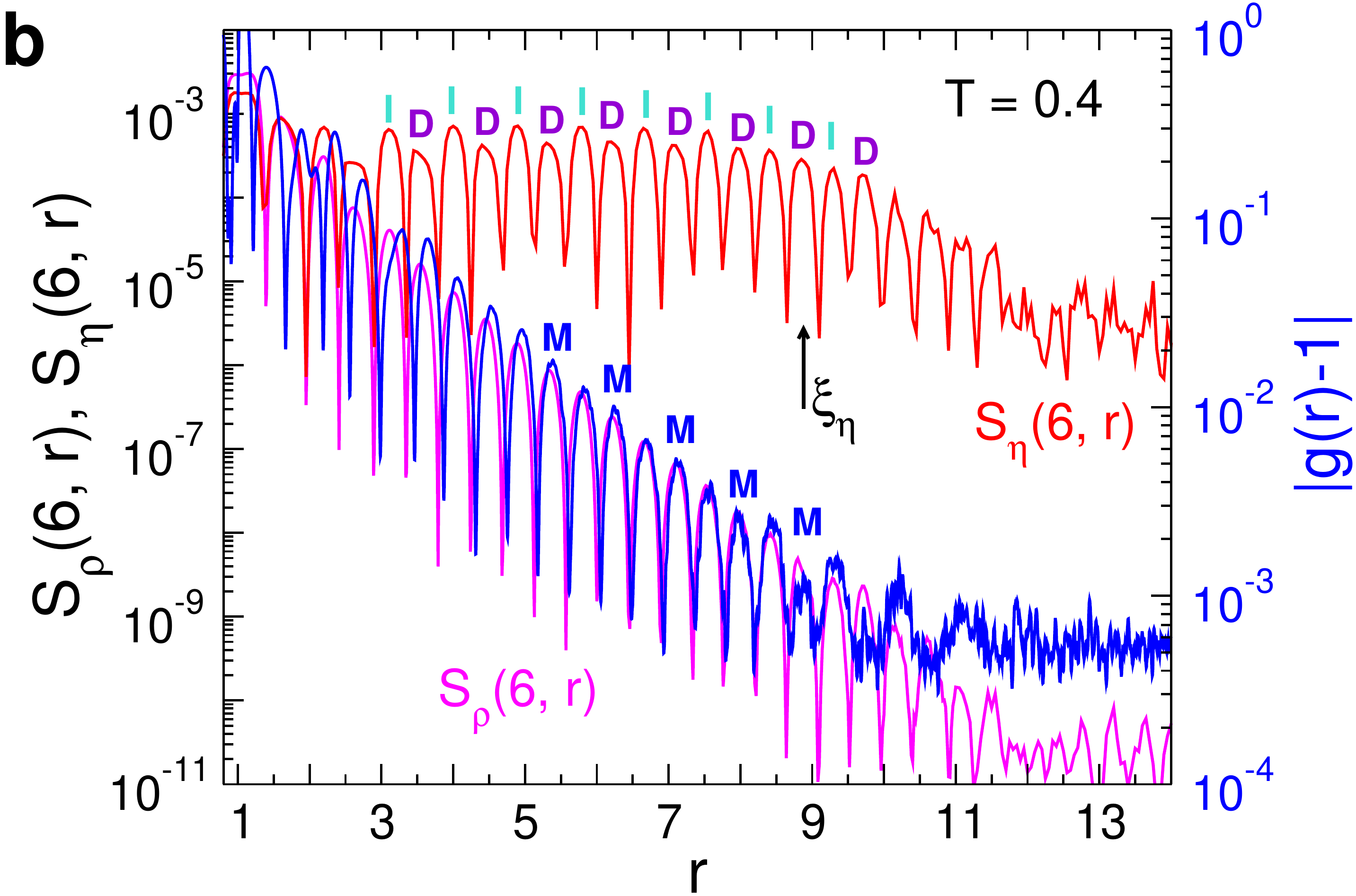}
\includegraphics[width=0.5\columnwidth]{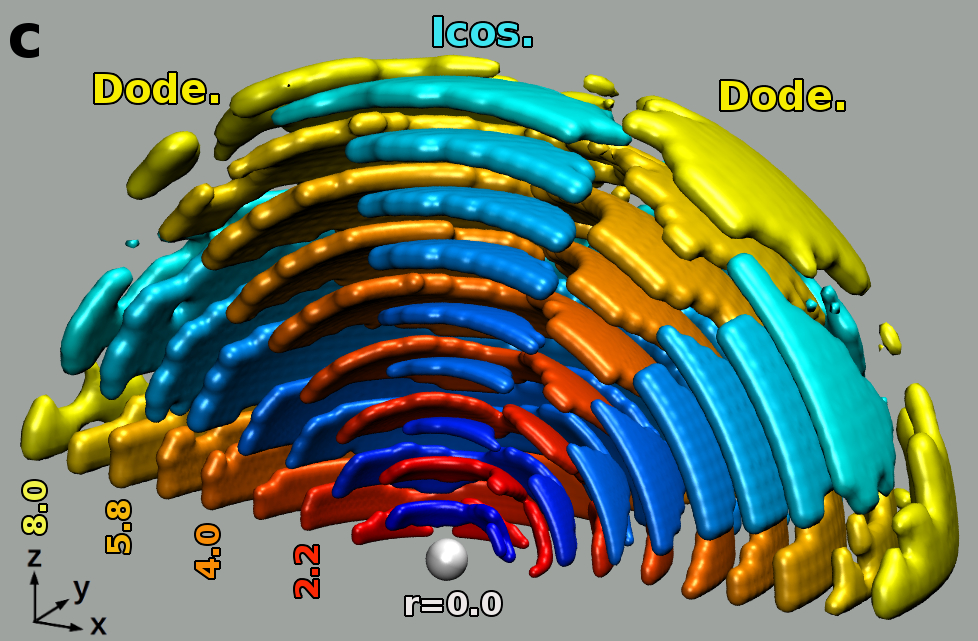}
\caption{
{\bf Quantitative characterization of the structural order.} {\bf a}
and {\bf b}: The angular power spectra and radial distribution function
for the liquids at $T=2.0$, {\bf a}, and $T=0.4$, {\bf b}.  The power spectrum
$S_\rho(6,r)$ (magenta curve) shows an exponential-like decay as a function
of the distance $r$. The power spectrum for the normalized density
distribution, $S_\eta(6,r)$ (red curve), stays large even at intermediate
$r$. $S_\eta(r)$ starts to decrease if $r$ is beyond a $T-$dependent
threshold, indicating the presence of a static correlation length. For
$r \gtrsim 4.0$ the high/low maxima in $S_\eta(r)$, labeled I and D,
coincide with the minima/maxima (labeled M) in $|g(r)-1|$ (blue line,
right ordinate). This up-down behavior is related to the alternating
icosahedral/dodecahedral symmetry in the distribution of the particles
when $r$ is increased. Note that the abscissa in the two panels have
different scales. {\bf c}: Three dimensional representation of the
layered structure extending to large distances for $T=0.4$. Only regions
with high density (covering 35\% area of the sphere) are depicted. The
bluish/reddish colors correspond to the locations of the high/low maxima
in $S_\eta(r)$ and thus to shells with icosahedral/dodecahedral symmetry.
}
\label{fig2_sff_gr}
\end{figure}

Figure~\ref{fig1_densityplot}{\bf c}-{\bf l} shows the three dimensional
normalized distribution $\rho(\theta,\phi,r)$ of the particles on the
sphere of radius $r$ centered at a particle \#1.  We recognize that
$\rho(\theta,\phi,r)$ has a noticeable angular dependence not only at
small distances but also at intermediate ones, e.g.~at $r=4.5$, and at low
$T$ even at large ones, e.g.,~$r=8.0$ (Fig.~\ref{fig1_densityplot}{\bf l}),
demonstrating that the liquid has a non-trivial structural order
that extents to distances that are well beyond the first few
neighbor shells. Furthermore one notes that $\rho(\theta,\phi,r)$ has
a highly symmetric shape: For distances $r\approx1$, i.e.,~the
first nearest neighbor shell, one finds the expected icosahedral
symmetry~\cite{jonsson_88,malins_13}, see~\ref{SI_fig_1stshell}.
For $r=1.65$ (Fig.~\ref{fig2_sff_gr}, {\bf c} and {\bf h}), corresponding to the
distance between the first minimum and the second nearest neighbor peak
in $g(r)$, one observes a dodecahedral-like symmetry. This result can be
understood by recalling that a dodecahedron is the dual of an icosahedron,
and vice versa (Fig.~\ref{fig1_densityplot}{\bf b}) and hence the local dip
formed by three neighboring particles in the first shell will be occupied
by particles forming part of the second shell, thus giving rise to a
dodecahedral symmetry. The fact that this ``duality mechanism'' works
even at large distances, see below, is surprising since it contradicts
the standard view that in liquids correlations are quickly washed
out at large distances. We emphasize that for geometrical reasons at
large $r$ a region with high $\rho(\theta,\phi,r)$ is {\it not} a single
particle, but a structure that grows linearly with $r$ and hence is a
whole collection of particles, i.e., for fixed $r$ the structure
is given by patches with a high density of particles that alternate with
patches with low density.

The standard way to characterize in a quantitative manner the density
distribution on a sphere is to decompose it
into spherical harmonics $Y_l^m$, $\rho(\theta,\phi,r) =$\linebreak
$\sum_{l=0}^\infty \sum_{m=-l}^{l}\rho_l^m(r) Y_l^m(\theta,\phi)$,
where the expansion coefficients $\rho_l^m$ are given in the
Methods, and to consider the angular power spectrum $S_\rho(l,r)=
(2l+1)^{-1}\sum_{m=-l}^{l}|\rho_l^m(r)|^2$. For this system the component
with $l=6$ is the most prominent one (\ref{SI_fig_sff_S_compare-l}{\bf a}),
independent of $r$, a result that is reasonable in view of the icosahedral
and dodecahedral symmetries that we find in the density distribution.
(For systems like SiO$_2$ which have a local tetrahedral symmetry the
relevant index is instead $l=3$.) We emphasize, however, that the results
presented below do not change qualitatively if another value of $l$
is considered (\ref{SI_fig_sff_S_compare-l}{\bf b}).

In Fig.~\ref{fig2_sff_gr} we show the $r-$dependence of $S_\rho(6,r)$
at a high and low temperature and one sees that the signal decays
quickly with increasing $r$. Figure~\ref{fig1_densityplot}{\bf k} shows
that for the distance $r=5.85$ the density distribution has a pronounced
structure although Fig.~\ref{fig2_sff_gr}{\bf b} shows that at this $r$
the absolute value of $S_{\rho}(l,r)$ is small. This smallness is
due to the fact that $S_{\rho}(l,r)$ is not only sensitive to the
angular dependence of the distribution, but also to the amplitude of
the signal. In order to probe the symmetry properties of the density
distribution it is therefore useful to consider a {\it normalized} density
distribution $\eta(\theta,\phi,r)=[\rho(\theta,\phi,r)-\rho_{\rm min}(r)]/[\rho_{\rm
max}(r)-\rho_{\rm min}(r)]$, where $\rho_{\rm max}(r)$ and $\rho_{\rm min}(r)$ are the
maximum and minimum of $\rho(\theta,\phi,r)$, respectively. The angular
power spectrum of $\eta(\theta,\phi,r)$, $S_{\eta}(r)$, is included in
Fig.~\ref{fig2_sff_gr}{\bf a}/{\bf b} as well and we see that this quantity oscillates
around a constant value which shows that the density distribution
has a pronounced angular dependence even at intermediate distances.
For distances larger than a threshold $\xi_{\eta}(T)$, $S_{\eta}(r)$
starts to decay before it reaches at large $r$ a value that is determined
by the noise of the data and below we will discuss the $T-$dependence
of $\xi_{\eta}(T)$. (See Methods for a precise definition of $\xi_\eta$.)

Most remarkable is the observation that for distances larger than
$r \approx 2.0$ the height of the local maxima in $S_\eta(r)$ shows
a periodic behavior in that a high maximum is followed by a low
one. A visual inspection of $\rho(\theta,\phi,r)$ reveals that these
high/low maxima correspond to distances at which the distribution
has an icosahedral/dodecahedral symmetry demonstrating that these two
geometries are not only present at short distances but also at large
ones, in agreement with the snapshots in Fig.~\ref{fig1_densityplot}.
One thus concludes that the distribution of the
particles in three dimensions is given by shells in which the particles
are arranged in a pattern with alternating icosahedral/dodecahedral
symmetry, see Fig.~\ref{fig2_sff_gr}{\bf c}. For distances larger that
$r\approx 4$ one finds that the radial position of these two geometrical
arrangements match well the locations of the minima/maxima in $g(r)$
(Fig.~\ref{fig2_sff_gr}, {\bf a} and {\bf b}). This observation
can be rationalized by the fact that a dodecahedron has 20 vertices
(i.e., regions in which $\rho(\theta,\phi,r)$ has high values) and an
icosahedron only 12, thus making that the former structure corresponds
to the {\it maxima} of $g(r)$ and the latter to the {\it minima}. In
contrast to this one finds no noticeable correspondence between the
peaks in $S_\eta(r)$ and $g(r)$ for $r < 3$, see also \ref{SI_fig_S_small_r},
indicating that the packing in the first few shells around the central
particle has not just a pure icosahedral or dodecahedral symmetry but
a more complex structure that is determined by steric and energetic
considerations, a result that is in agreement with previous studies of
similar systems that have probed the geometry of the packing on small
length scales~\cite{coslovich_07,royall_15,ma_11,miracle_04}.

\begin{figure}[tb]
\center
\includegraphics[width=0.65\columnwidth]{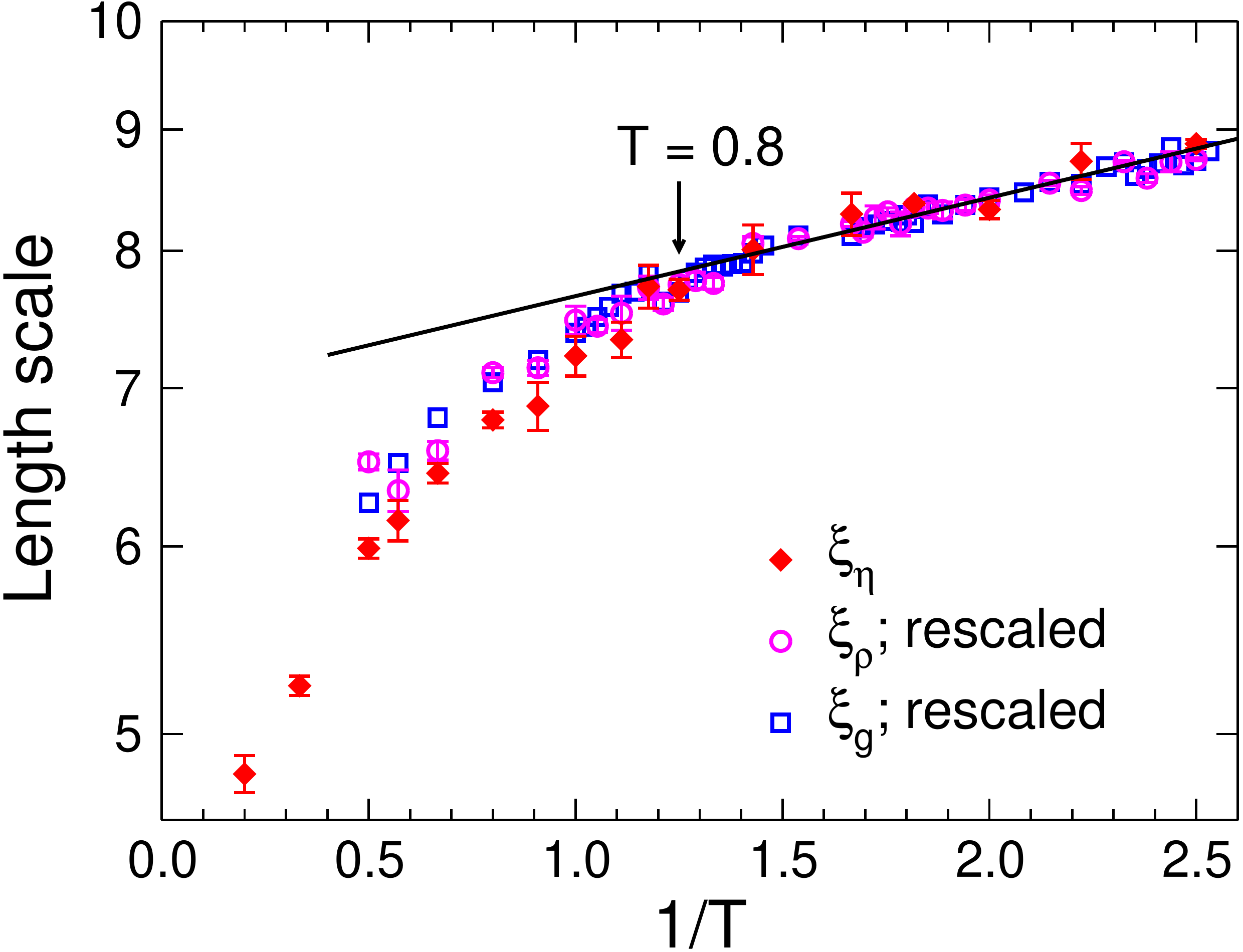}
\caption{
{\bf Temperature dependence of length scales.}
Different length scales (on log scale) as a function of inverse
temperature: $\xi_\eta$ defined from $S_\eta(r)$ is shown in red
and the inverse of the slope of the exponential decay of $S_\rho(r)$
and $|g(r)-1)|$ is shown in magenta and blue, respectively. $\xi_\rho$
and $\xi_g$ have been multiplied by a scaling factor of 6.19 and 1.30,
respectively. The line is a guide to the eye to allow to identify the two
temperature dependencies that join at the cross-over temperature around
$T=0.8$. Error bars were obtained from the fits mentioned in the Methods.
}
\label{fig3_slopes}
\end{figure}

The distance $\xi_\eta(T)$ at which $S_\eta(r)$ starts to drop,
see Fig.~\ref{fig2_sff_gr}{\bf a}/{\bf b}, corresponds to a static
correlation length (\ref{SI_fig_peak}).
Figure~\ref{fig3_slopes} shows $\xi_\eta$ as a function of inverse
temperature and one recognizes that this length increases by about a
factor of two in the $T-$range considered. Also included in the graph are
the length scales $\xi_\rho$ and $\xi_g$ that are related to the exponential decay of
$S_\rho(r)$ and $|g(r)-1|$, respectively, see Fig.~\ref{fig2_sff_gr}{\bf
a}/{\bf b}. (See the Methods and \ref{SI_fig_peak} on how to determine
$\xi_\rho$ and $\xi_g$.)  From the graph one recognizes two regimes: At
high $T$ the length scales increase quickly with decreasing $T$ whereas at low
temperatures one finds a weaker $T-$dependence and which is compatible
with $\ln(\xi) \propto T^{-1}$.  Hence one concludes that a decreasing
temperature leads to an increasing static length scale, in agreement
with previous studies that have documented a weak increase of static
length scales in glass-forming systems, Ref.~\cite{royall_17} and references
therein. Surprisingly the crossover between the two regimes occurs at
around $T=0.8$, thus very close to the so-called ``onset temperature''
$T_{\rm o}$~\cite{kob_95} at which the relaxation dynamics of the system
crosses over from a normal dynamics to a glassy one~\cite{binder_11}. This
result shows that the change in the dynamical properties of the system
has a counterpart in the statics, giving hence support to the idea that
the latter allows to understand the former~\cite{mct}.

Since the $T$-dependence of $\xi_g$ is very similar to the one of
$\xi_\eta$, Fig.~\ref{fig3_slopes}, one can expect that also the intensity
of the static structure factor $S(q)$ at small wave-vectors has the same
$T-$dependence. \ref{SI_fig_sq}{\bf d} shows that this is indeed the case
(and the same conclusion is reached for the compressibility) which thus
makes this $T-$dependence accessible to standard scattering experiments,
i.e., a careful measurement of the structure factor allows to determine
the $T$-dependence of the length scale $\xi_\eta$ as well as the onset
temperature in a direct manner, i.e., without referring to any probe
of the {\it dynamics}. 

In conclusion we have demonstrated that liquids have a non-trivial
structural correlation that extents to distances well beyond the first
few nearest neighbors. This result has been obtained by using a novel
method to analyze the particle coordinates and which can thus
be applied to any other disordered system for which the particle
coordinates are accessible, such as colloidal and granular systems, or
materials in which some of the particles have been marked by fluorescence
techniques~\cite{xia_17,kegel_00,weeks_00,sherson_10,kou_17}. Our finding
that disordered systems can have anisotropic structural order extending
to large length scales should trigger the improvement of experimental
techniques that probe this order.


\section*{Acknowledgments}
 We thank D. Coslovich, G. Monaco, M. Ozawa, and K. Schweizer for discussions. 
{\bf {Funding:}}
Part of this work was supported by the China Scholarship Council grant 201606050112 
and grant ANR-15-CE30-0003-02.

{\bf {Author contributions:}}
Z.Z. and W.K. designed the research and carried out the simulations. Z.Z. analyzed the data. Z.Z. and W.K. wrote the paper.
{\bf {Competing interests:}} The authors declare no competing financial interests. 
{\bf {Data and materials availability:}} All data in the manuscript or
the Materials are available from W. Kob upon reasonable request.

\renewcommand{\figurename}{\bf Fig.}

\clearpage

\setcounter{page}{1}

\clearpage
\section*{Methods}
{\bf System and simulations.} 
The system we study is a 
80:20 mixture of Lennard-Jones particles (type A and B) with
interactions given by $V_{\alpha\beta}(r)=4\epsilon_{\alpha\beta}
[(\sigma_{\alpha\beta}/r)^{12}-(\sigma_{\alpha\beta}/r)^{6}]$, where
$\alpha,\beta \in \{A,B\}$, $\sigma_{AA}=1.0$, $\epsilon_{AA}=1.0$,
$\sigma_{AB}=0.8$, $\epsilon_{AB}=1.5$, $\sigma_{BB}=0.88$, and
$\epsilon_{BB}=0.5$~\cite{kob_95}. Here we use $\sigma_{AA}$ and
$\epsilon_{AA}$ as the units of length and energy, respectively. We set
the mass of all particles equal to $m=1.0$ and the Boltzmann constant
is $k_B=1.0$. Using the LAMMPS software~$^{30}$ we simulate
a total of $10^5$ particles at constant volume (box size 43.68) and
temperature. At the lowest temperature, $T=0.40$, the run was $1.4\cdot
10^8$ time steps (step size 0.005) for equilibration and the same length
for production, time spans that are sufficiently large to completely
equilibrate the system. For the analysis of the data we used 4 and 20
configuration for $S_\rho$ and $g(r)$, respectively.\\

\noindent
{\bf Angular power spectrum.}
The coefficient $\rho_l^m$ for the expansion of the density distribution into
spherical harmonics is given by
\begin{equation}
\rho_l^m =\int_0^{2\pi} d\phi \int_0^\pi d\theta \sin \theta  
\rho(\theta,\phi,r) Y_l^{m*}(\theta,\phi) \quad ,
\nonumber
\end{equation}

\noindent
where $Y_l^{m*}$ is the complex conjugate of the spherical harmonic
function of degree $l$ and order $m$.  In practice this integration was
done by sampling the integrand over up to $2\cdot 10^9$ points for each
shell of width 0.4.\\

\renewcommand{\figurename}{}
\renewcommand{\thefigure}{Extended Data Fig. \arabic{figure}}
\setcounter{figure}{0}

\begin{figure}[ht]
\center
\includegraphics[width=0.45\columnwidth]{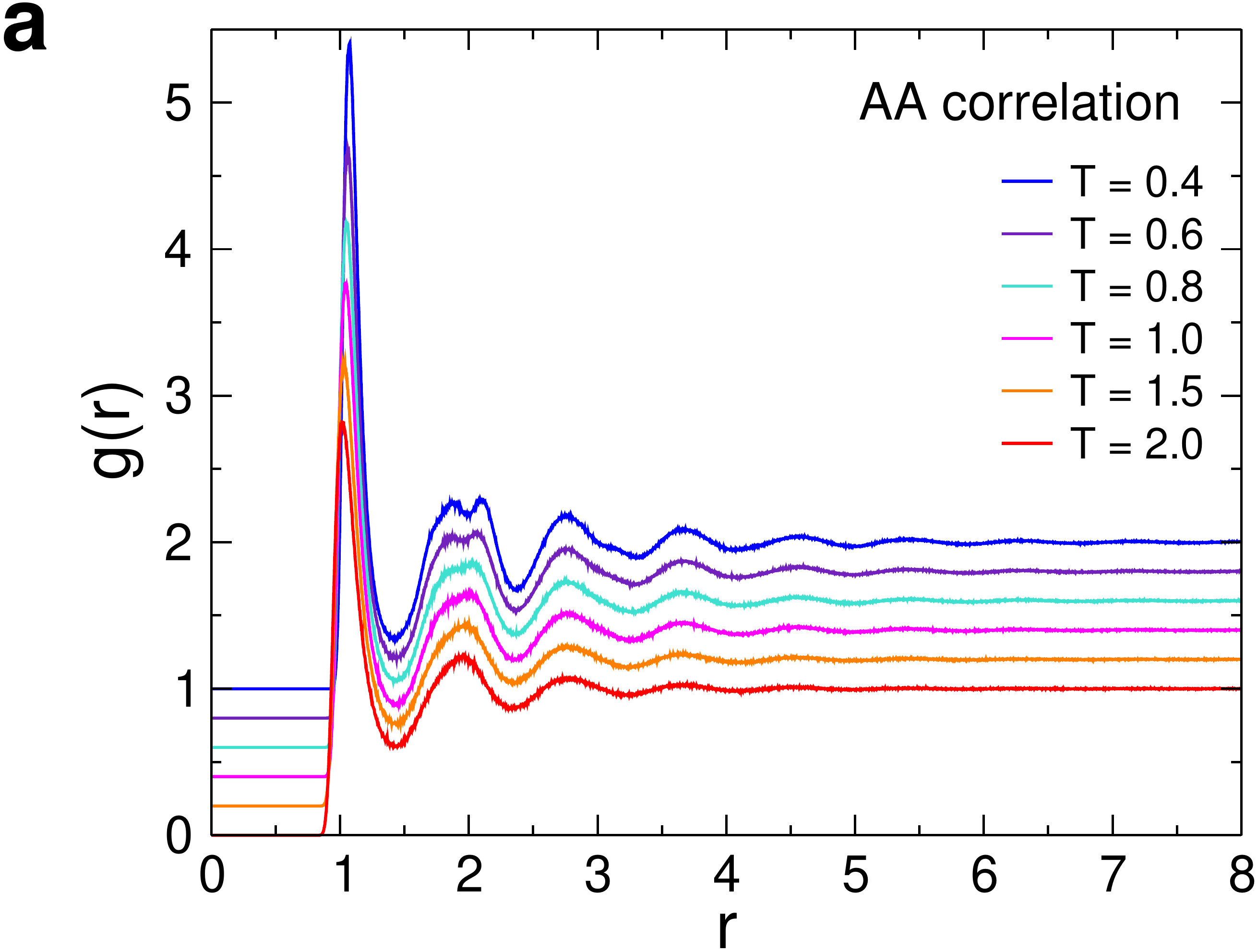}
\includegraphics[width=0.45\columnwidth]{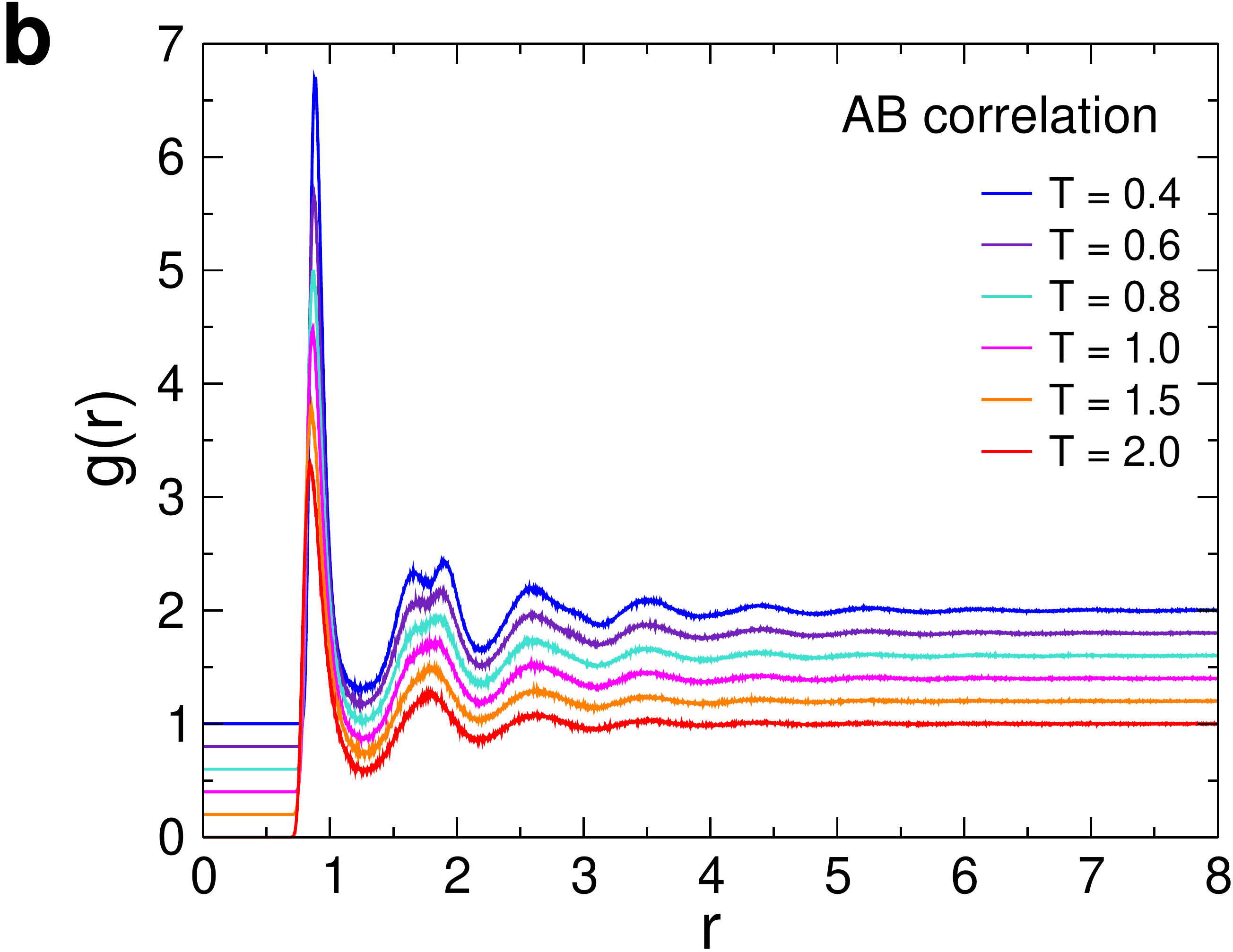}
\includegraphics[width=0.45\columnwidth]{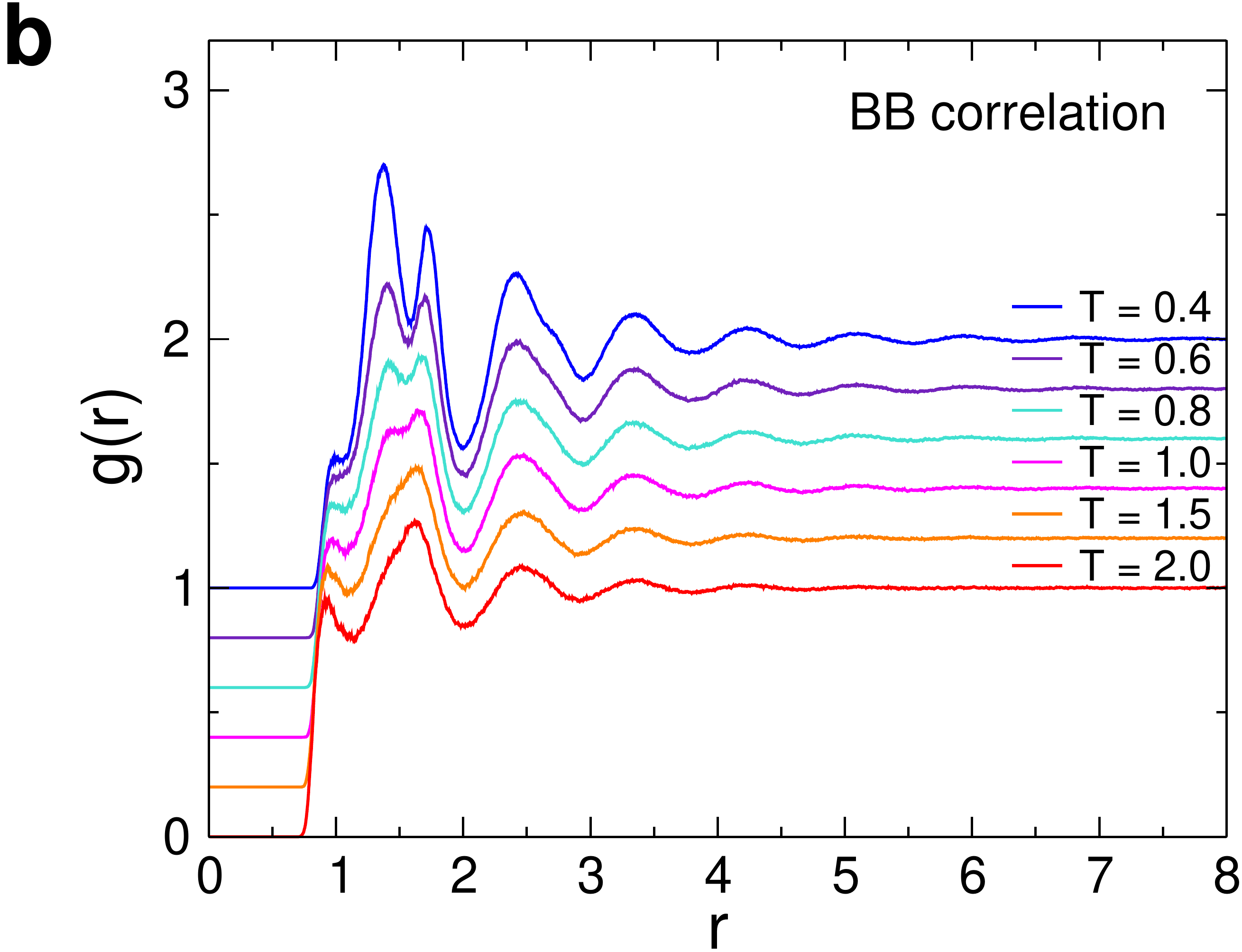}
\includegraphics[width=0.45\columnwidth]{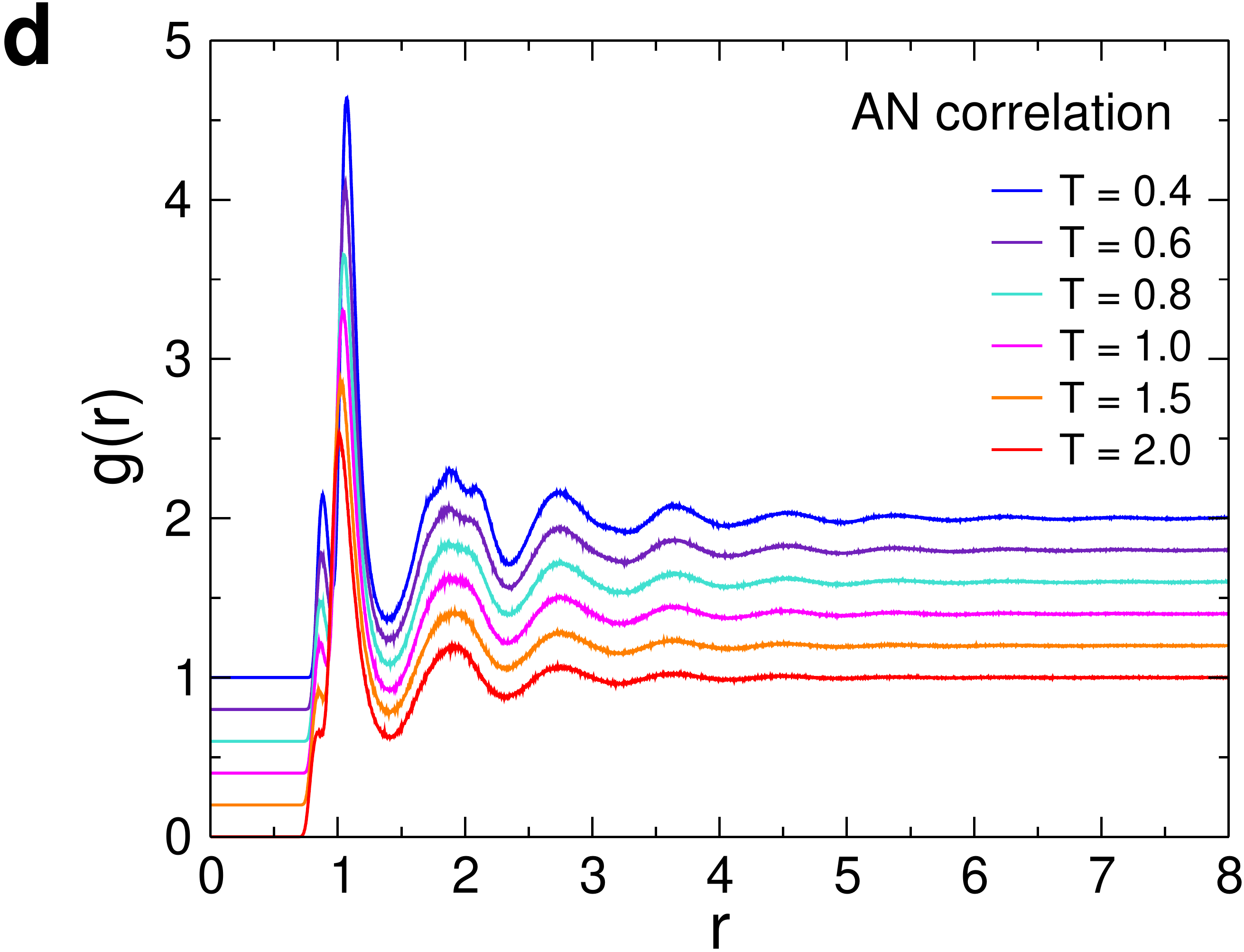}
\caption{{\bf Radial distribution functions for several temperatures.} 
The AA, AB, BB, and AN correlations are shown
in panels {\bf a}, {\bf b}, {\bf c}, and {\bf d}, respectively. For the sake of clarity the different curves
have been shifted vertically by multiples of 0.2.
}
\label{SI_fig_gofr}
\end{figure}

\noindent
{\bf Radial distribution functions.}
In \ref{SI_fig_gofr} we show
the three partial radial distribution functions $g(r)$ as well as the
correlation function between an A particle and any other particle (AN).
The curves correspond to different temperatures (see legend).  These graphs show
that the $T-$dependence of $g(r)$ is very smooth, as expected for a
system that is a good glass-former.\\

\begin{figure}[ht]
\center
\includegraphics[width=0.95\columnwidth]{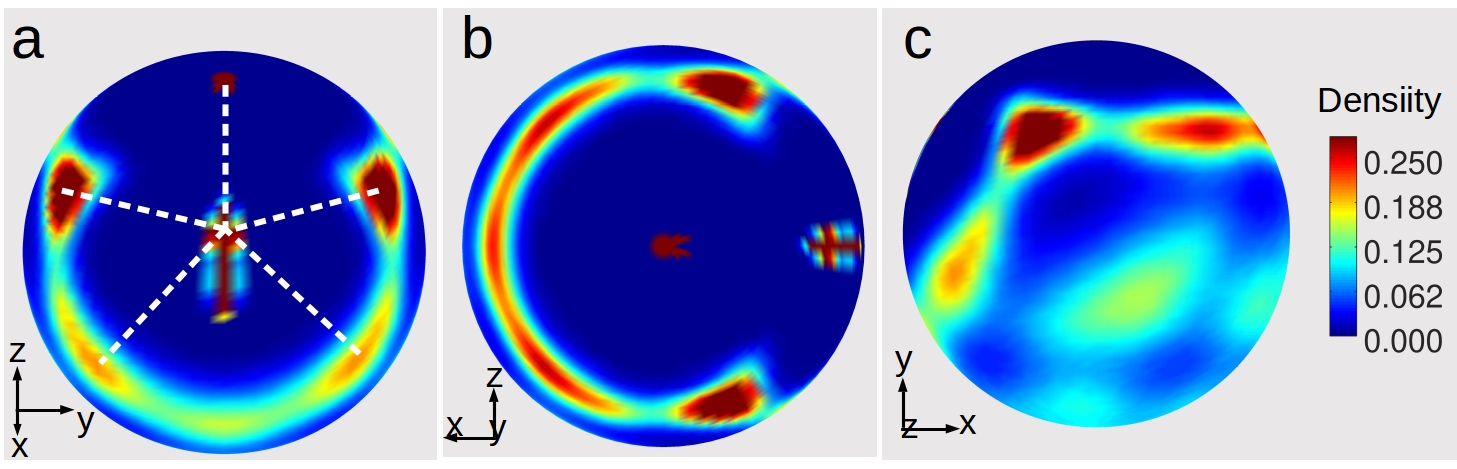}
\caption{{\bf Three dimensional distribution of the particles in the nearest neighbor shell.}
Density distribution $\rho(\theta,\phi,r)$ for $r=1.1$, i.e.~the
distance that corresponds to the particles in the first shell of the
central particle. The temperature is $T=0.4$. {\bf a} to {\bf c}:
Different perspectives of the density distribution on the sphere
(see orientation of the coordinate system). A pronounced icosahedral-like
symmetry can be recognized, see, e.g., the view of the distribution along
the $x$-axis in panel {\bf a} where the dashed lines indicate the connection
between neighboring particles.
}
\label{SI_fig_1stshell}
\end{figure}

\noindent
{\bf Three dimensional distribution of the particles in the nearest neighbor shell.}
In Fig.~\ref{fig1_densityplot} of the main text we show the angular
distribution of the particle density for intermediate and large distances $r$.
In \ref{SI_fig_1stshell} we present this distribution for a
distance that corresponds to the first coordination shell of the central particle.
This graph clearly shows that this first shell has an icosahedral-like
symmetry, as expected for a hard-sphere like simple liquid.\\

\begin{figure}[ht]
\center
\includegraphics[width=0.60\columnwidth]{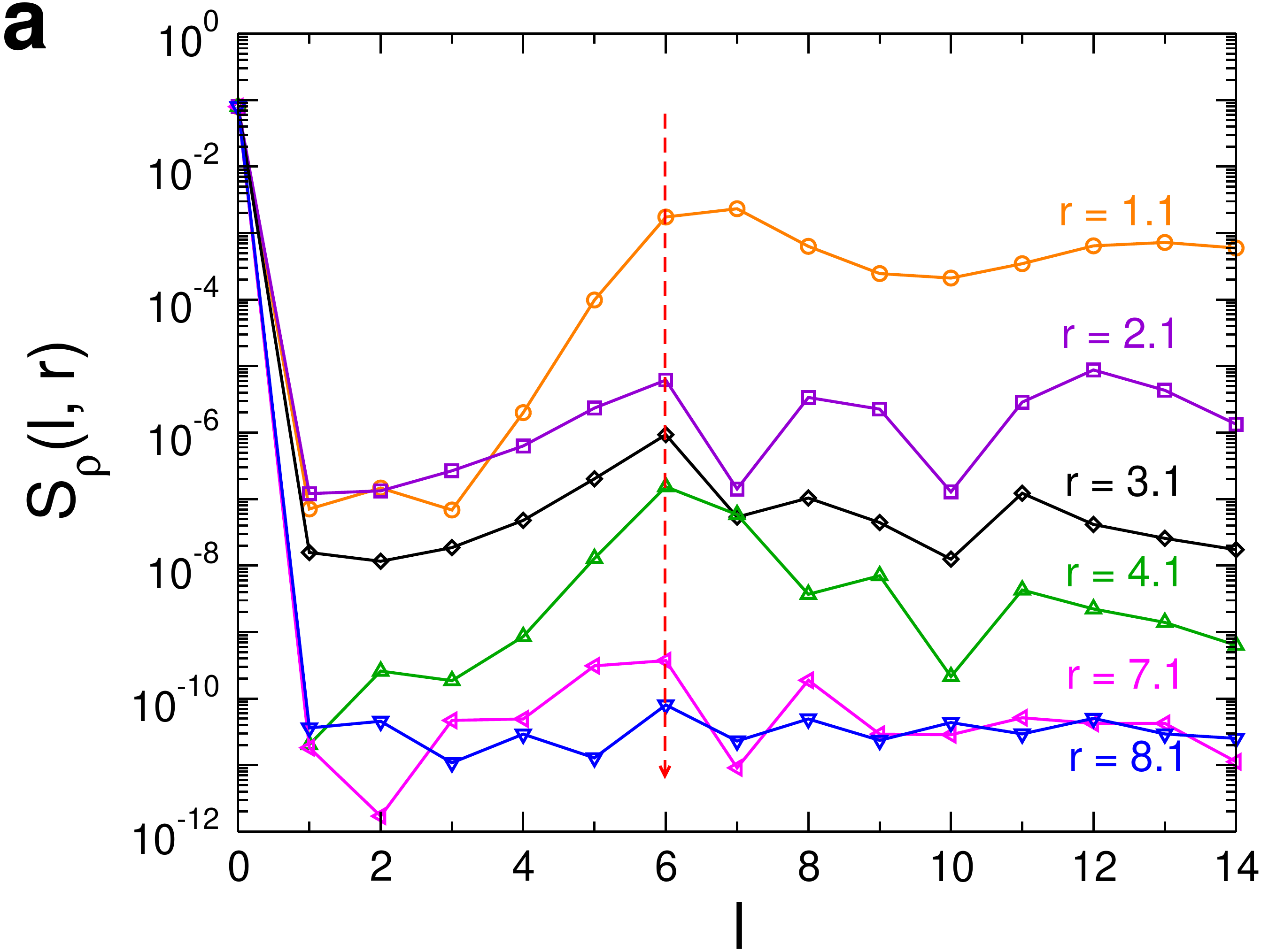}
\includegraphics[width=0.60\columnwidth]{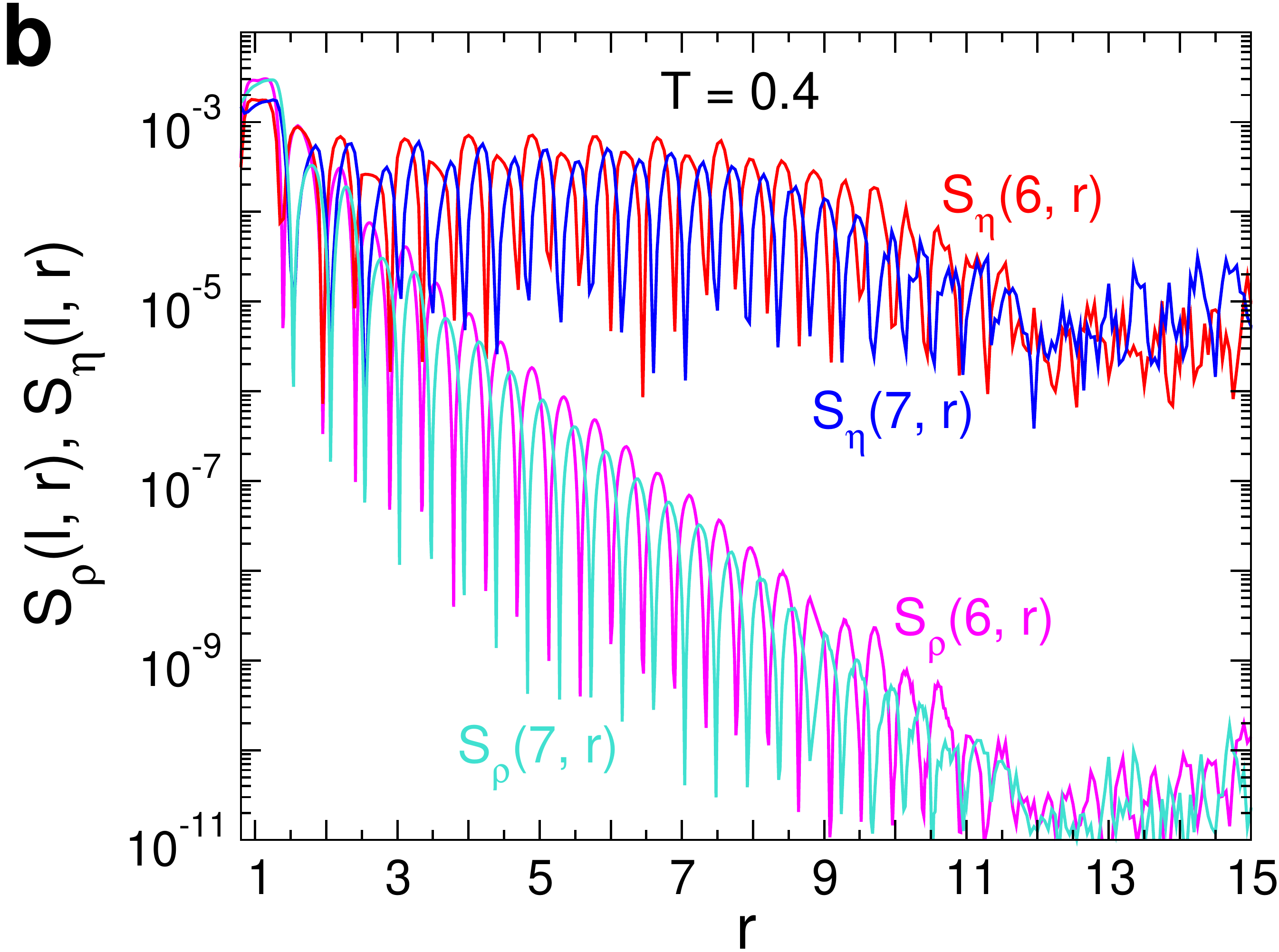}
\caption{{\bf $l-$dependence of the angular power spectrum.}
{\bf a}: $S_\rho(l,r)$ for $T=0.5$. The curves correspond to the indicated values
of $r$. One sees that the signal for $l=6$, marked by a vertical dashed line, 
is pronounced at all distances.
{\bf b}: Comparison between the angular power spectra for $l = 6$ and $l =
7$. The $r-$dependence of $S_\rho(l,r)$ and $S_\eta(l,r)$
are qualitatively the same.
}
\label{SI_fig_sff_S_compare-l}
\end{figure}

\noindent
{\bf $l-$dependence of the angular power spectrum.}
In the main text we focus on the results for the index $l=6$ in
the expansion of the spherical harmonics of the density distribution. In
\ref{SI_fig_sff_S_compare-l}{\bf a} we show the $l-$dependence of the angular
power spectrum $S_\rho(l,r)$ for selected distances $r$. From this graph one
recognizes that for $l=6$ the signal is relatively strong for {\it all}
distances and hence this value for the index is a good choice for probing
the structural order in the liquid. However, as mentioned in the main
text, the presented results do not depend in a crucial manner on the
choice of $l$. This is demonstrated in \ref{SI_fig_sff_S_compare-l}{\bf b}
where we compare the $r-$dependence of $S_\rho$ and $S_\eta$ for $l=7$
with the ones for $l=6$, i.e., the data shown in the main text. This figure
clearly demonstrates that the $r-$dependence of the two quantities does
not depend in a significant manner on $l$. (The main difference is that
the signal for $l=7$ is somewhat smaller than the one for $l=6$, a result
that is directly related to the fact that in this system the particles
are arranged in shells with a pronounced icosahedral and dodecahedral
symmetry and that these two structures have a relatively strong $l=6$
component in the angular power spectrum.) Therefore the corresponding
length scales (see Fig.~\ref{fig3_slopes} in the main text), the radial
distribution function, as well as the signal of the static structure
factor at small wave-vector will all show a similar dependence on
temperature (see also \ref{SI_fig_sq} below).  \\

\begin{figure}[ht]
\includegraphics[width=0.8\columnwidth]{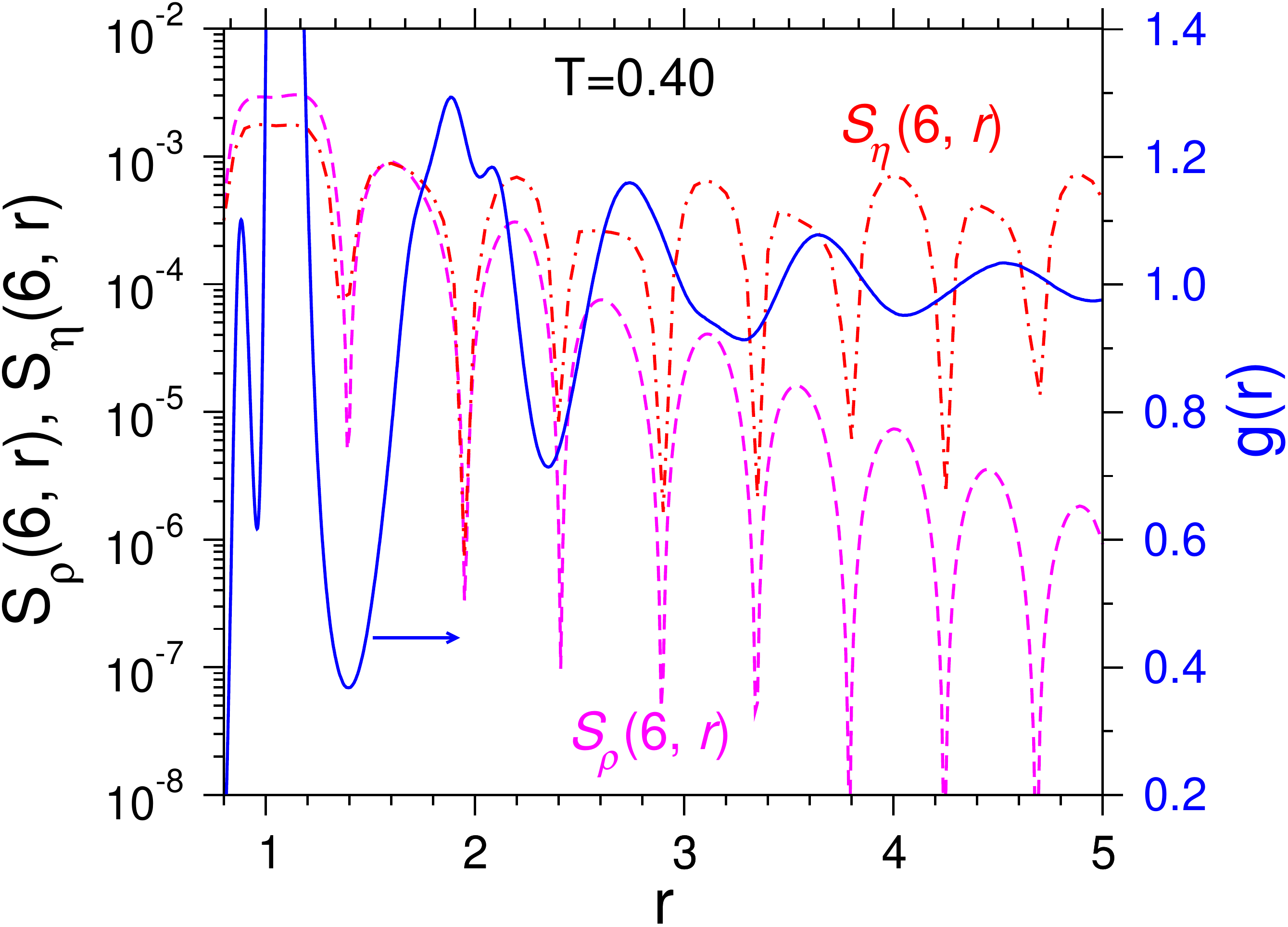}
\caption{{\bf Angular power spectra and radial distribution function at short distances.}
$T = 0.4$ and $l=6$.  Note that the double peaks in the first shell, i.e. $r
\approx 1.0$ originate from A-B (smaller peak) and A-A (bigger peak)
correlations (see Fig.~\ref{SI_fig_gofr}).
}
\label{SI_fig_S_small_r}
\end{figure}

\noindent
{\bf Angular power spectra and radial distribution function at short distances.}
In Fig.~\ref{fig2_sff_gr} of the main text we have shown how the angular
power spectra $S_\rho(r)$ and $S_\eta(r)$ and the radial distribution
function $g(r)$ depend on the distance $r$. In \ref{SI_fig_S_small_r} we
show these functions at small distances, i.e., $r<5.0$. One recognizes
from this graph that at these small distances, in particular for
$r<3.0$, the $r-$dependence is rather complex due to the local packing
effects of the particles. Only for distances larger than around 4.0 the
$r-$dependence of the three functions becomes regular in that the shape
of the various peaks becomes independent of $r$. Thus this distance
indicates the crossover between a structure at small $r$ that is
determined by local packing effects to a structure at large $r$ that
is determined by symmetry considerations.  This symmetry at large $r$ is in turn
determined by the packing at {\it small r}, i.e., in
our case the icosahedra-like structure.\\

\begin{figure}[ht]
\includegraphics[width=0.45\columnwidth]{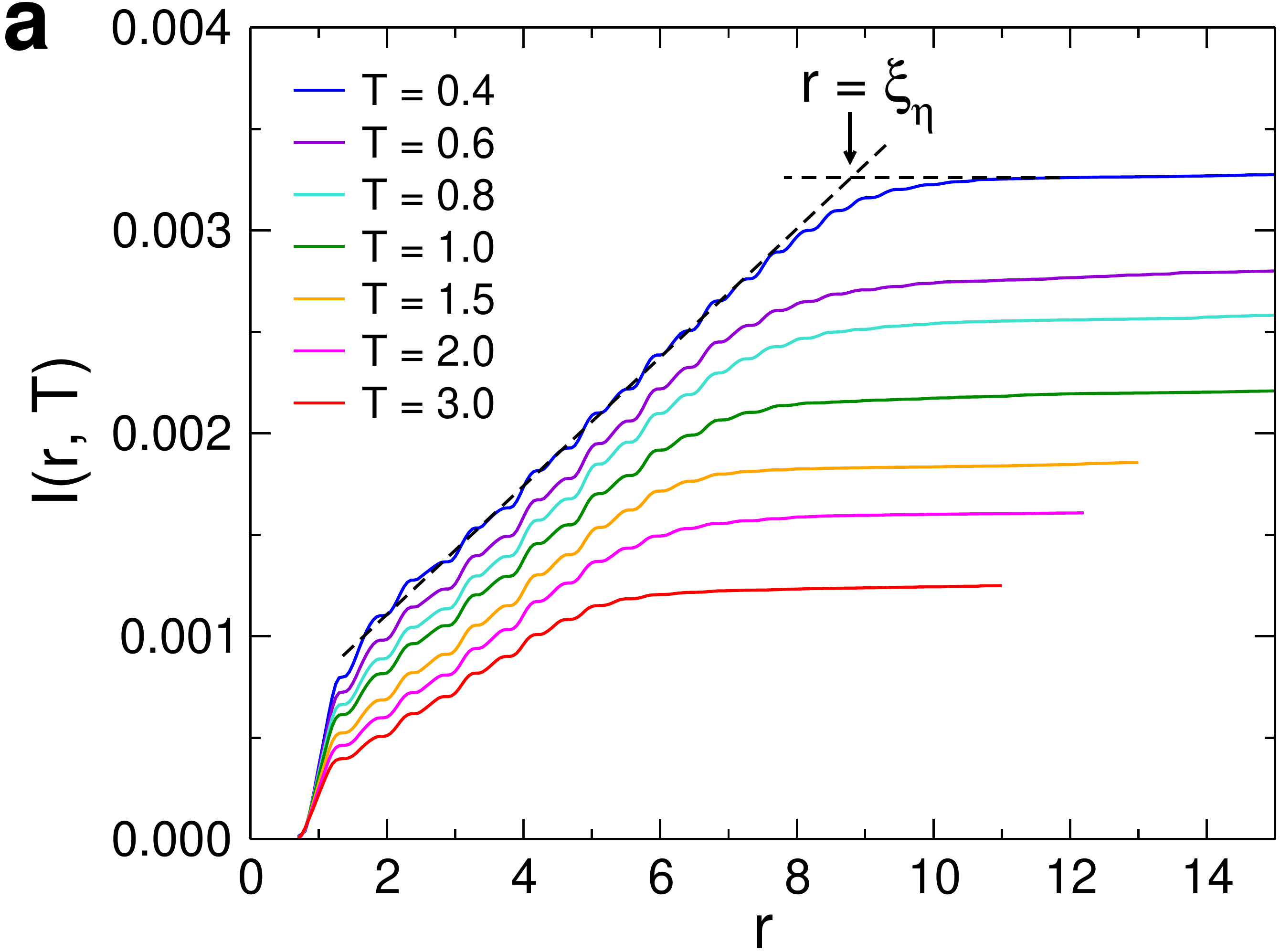}
\includegraphics[width=0.45\columnwidth]{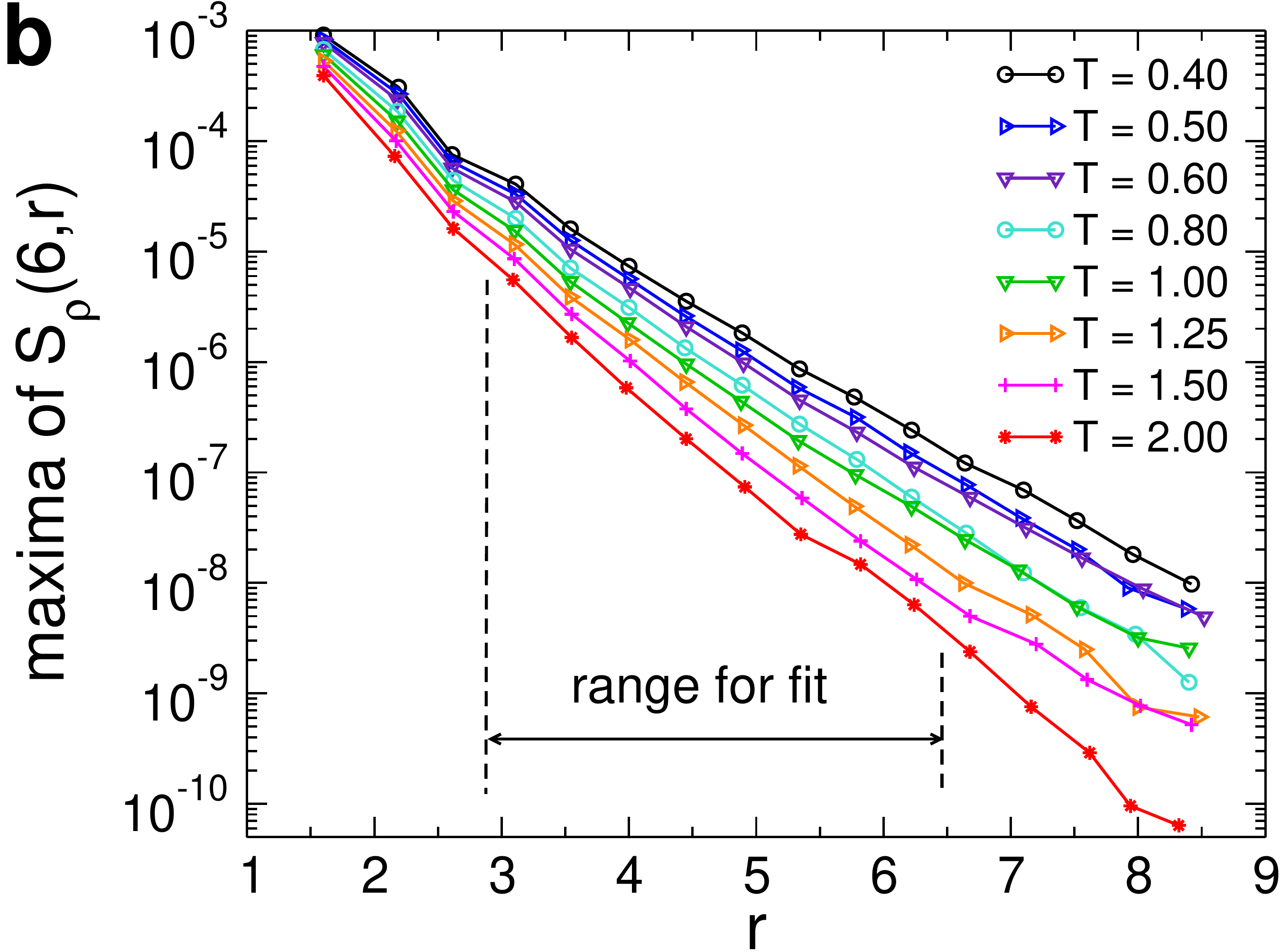}
\includegraphics[width=0.45\columnwidth]{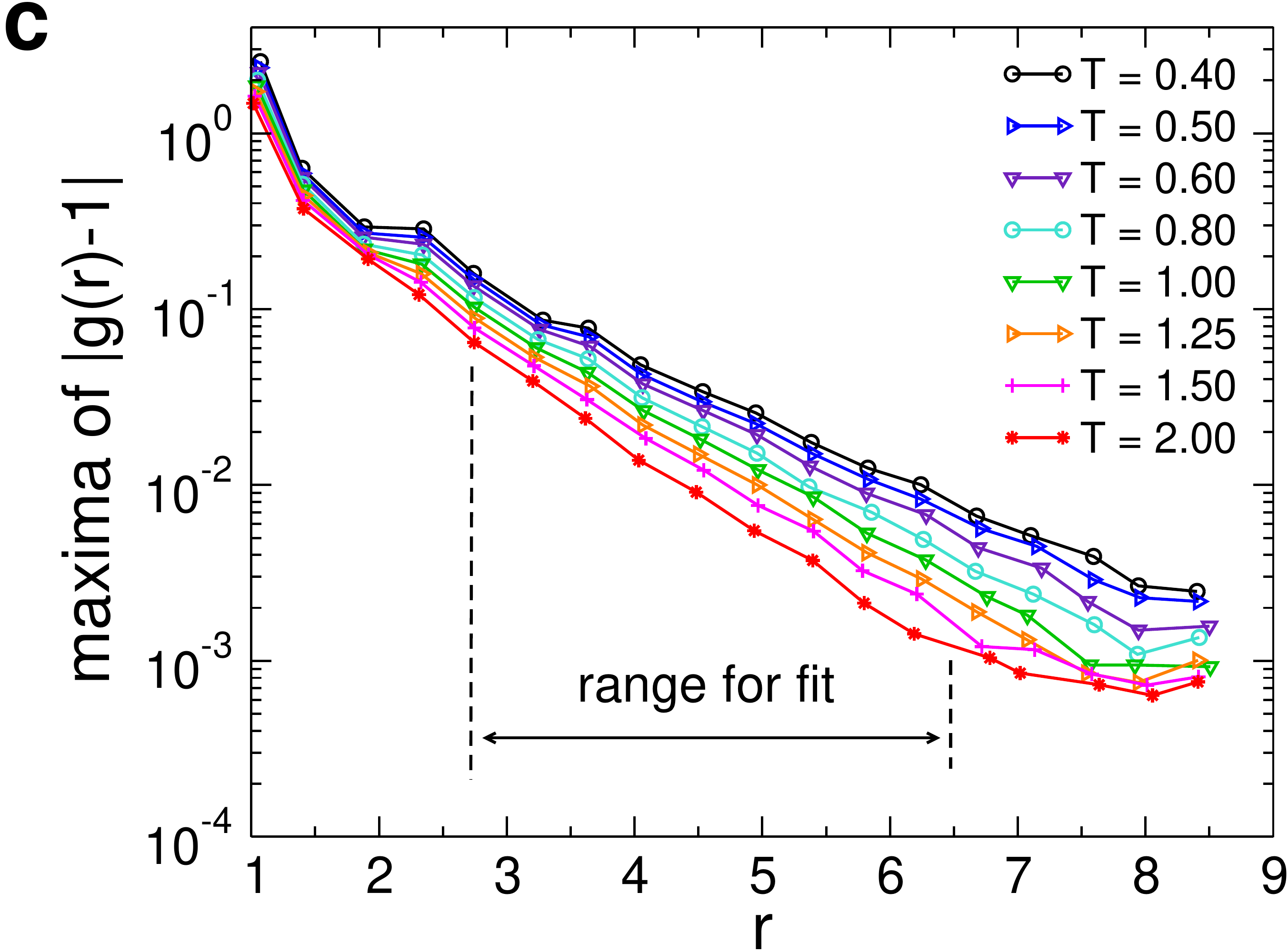}
\caption{{\bf Extracting length scales.}
{\bf a}: $I(r,T)$, the integral of $S_\eta(r)$ for different temperatures.
The length scale $\xi_\eta(T)$ is defined as the crossover point at
which $I(r,T)$ starts to become a constant (see dashed lines).
{\bf b}: Local maxima of $S_\rho(6,r)$. {\bf c}: Local maxima of
$|g(r)-1|$. For both quantities, the data in the range $2.8 < r < 6.5$
are fitted with an exponential function to extract the corresponding length scale.
}
\label{SI_fig_peak}
\end{figure}

\noindent
{\bf Extracting length scales.}
From Fig.~\ref{fig2_sff_gr}{\bf a}/{\bf b} of the main text one recognizes that the
distance $\xi_\eta$ at which $S_\eta(r,T)$ starts to drop increases if
$T$ is lowered. To determine $\xi_\eta$ we have calculated the integral
$I(r,T)=\int_0^r S_\eta(r',T)dr'$ and in \ref{SI_fig_peak}{\bf a} we
plot this quantity as a function of $r$. For small and intermediate $r$
the integral shows a basically linear increase with $r$, because the
integrand $S_\eta(r)$ is essentially a constant, and once $S_\eta(r)$
starts to decay $I(r,T)$ becomes a constant. Using a fit with two
straight lines this cross-over point can be determined accurately,
see dashed lines in \ref{SI_fig_peak}{\bf a}, giving thus $\xi_\eta(T)$.

In Fig.~\ref{fig2_sff_gr}{\bf a}/{\bf b} of the main text we have
shown that $S_\rho(r)$ shows at intermediate and large distances an
exponential dependence on the distance $r$. In \ref{SI_fig_peak}{\bf b}
we show the $r-$dependence of $S_\rho$ for different temperatures. Note
that we plot only the local maxima of the function since these have
been used to fit the data at intermediate and large distances with
an exponential function (see below). From the graph one recognizes
that the slope of the curves decreases with decreasing temperature,
indicating that the associated length scale increases. We obtain this
length scale $\xi_\rho$ by making a fit with an exponential of the form
$S_\rho(r,T) \propto \exp(-r/\xi_\rho(T))$ and plot this quantity in
Fig.~\ref{fig3_slopes} of the main text. In \ref{SI_fig_peak}{\bf c} we
show the $r-$dependence of $|g(r)-1|$ for various temperatures. (Again
only the location of the maxima are shown.) We see that also this
dependence can be fitted well by an exponential function, thus allowing
to define a length scale $\xi_g(T)$, the $T-$dependence of which is
included in Fig.~\ref{fig3_slopes} of the main text as well.\\

\noindent 
{\bf Structure factor and compressibility.}
The static structure factor of the system was determined directly from
the positions of the particles, i.e.,\\[-15mm]

\begin{equation}
S({\vec q}) =\frac{1}{N}\sum_{j=1}^N \sum_{k=1}^N\exp[{i\vec q}\cdot ({\vec r}_j-{\vec r}_k)]
\quad .
\label{SI_eq_sq}
\end{equation}

\noindent
Since the system is isotropic, we have averaged $S({\vec q})$ over
all wave-vectors ${\vec q}$ that have the same norm $q=|{\vec q}|$. In
\ref{SI_fig_sq}{\bf a} we show the $q-$dependence of $S(q)$
for different temperatures. Because of the finite size of the box,
the smallest accessible wave-vector is $q=2\pi/43.68\approx 0.144$
and one has only three independent wave-vectors with this modulus. In
order to estimate with good accuracy the $T-$dependence of $S(q)$ at
small wave-vectors we have therefore averaged $S(q)$ over the range $1.0 \leq q
\leq 2.0$. This interval is shown in \ref{SI_fig_sq}{\bf a} as well
(Inset).  The so obtained averaged data for $S(q)$, denoted by $S_0(T)$, is
shown in \ref{SI_fig_sq}{\bf c} (blue circles). In Fig.~\ref{fig3_slopes}
of the main text we have found that the various length scales show at
around $T=0.8$ a crossover in their temperature dependence. Since
this crossover is seen also in the $T-$dependence of the slope of
$|g(r)-1|$, one expects it to be present also in $S(q)$ at small $q$
and hence in $S_0(T)$. \ref{SI_fig_sq}{\bf c} includes also a
power-law fit to the low temperature data (solid blue line). (At this stage
this functional form should be considered just as a parameterization
of the data since we do not have a theoretical basis for it.)
In order to see better the $T-$range in which this fit works well,
we show in \ref{SI_fig_sq}{\bf d} the ratio between $S_0$ and
this power-law. One recognizes that this ratio shows an appreciable
$T-$dependence for $T\geq 0.85$, but then becomes flat, i.e.  the quantity
$S_0$ does indeed show a crossover at around the onset temperature
$T_{\rm o}\approx 0.8$. This result is thus coherent with the data shown
in Fig.~\ref{fig3_slopes} of the main text and hence we can conclude
that at the onset temperature the static structure of the system {\it
on large scales} is indeed changing its temperature dependence.

Furthermore we mention that we have also studied the temperature dependence
of $S_0(T)$ in the constant pressure ensemble. The chosen pressure was
$P=8.0$ since this corresponds to the pressure at the onset temperature
in the constant volume ensemble and hence it can be expected that the
onset temperature in the two ensemble are very similar.  The resulting
static structure factor is presented in \ref{SI_fig_sq}{\bf b} and
the so obtained $S_0(T)$ is included in \ref{SI_fig_sq}{\bf c} as
well (red squares). The data at low $T$ can again be fitted well by a
power-law (solid red line) and the resulting ratio between data and power-law, plotted
in \ref{SI_fig_sq}{\bf d}, shows again around $T=0.8$ a change in its
temperature dependence. This result demonstrates that the $T-$dependence of
the large scale structure shows at the onset temperature a marked change
which is independent of the considered ensemble. 

For the simulations at constant pressure we have determined also the
compressibility $\kappa= (\Delta V)^2/(k_B T V)$, where $(\Delta V)^2$
is the variance of the volume fluctuation. In \ref{SI_fig_sq}{\bf c}
we present thus also $\kappa(T)$ and we recognize that this quantity shows
a similar temperature dependence as $S_0$, as expected. Also in this
case we find that the data at low $T$ can be described very well by a
power-law (solid magenta line). In \ref{SI_fig_sq}{\bf d} we show this in
an explicit manner and one sees that the description with the power-law
starts to break down at around $T=0.85$, i.e., the same temperature at
which $S_0$ starts to deviate from the power-law. Thus one can conclude
that also a careful measurement of the compressibility does allow to
estimate the onset temperature $T_{\rm o}$ with good accuracy, or in
other words, this temperature that is usually obtained from {\it dynamic} data
can be extracted also from high precision {\it static} data.\\

\noindent
{\bf Anisotropic radial distribution function.}
In Fig.~\ref{fig2_sff_gr} of the main text and \ref{SI_fig_gofr}
we have presented the standard radial distribution function $g(r)$,
i.e., the density distribution averaged over the sphere with radius
$r$.  Since we find that the distribution of the particles around a
central particle is anisotropic it is of interest to consider also
the radial distribution functions in which one probes the correlations
in a specific direction with respect to the local coordinate system,
Fig.~\ref{fig1_densityplot}{\bf a}.  We have done this analysis for 
the directions that correspond to the vertices of the icosahedra and
of the dodecahedra, thus defining $g_I(r)$ and $g_D(r)$, respectively.
The resulting distribution functions are shown in \ref{SI_fig_gigd}. Panel
{\bf a} shows that, for intermediate and large distances, $g_D(r)$
has oscillations that are in phase with $g(r)$ whereas $g_I(r)$ has
oscillations that are in anti-phase. The amplitudes of the oscillations
in $g_I(r)$ and $g_D(r)$ are significantly larger than the ones found
in $g(r)$, a result that is reasonable since the latter function is
a weighted average of the two former ones and hence will be affected 
by cancellation effects.  Furthermore we have done the same analysis
also for the distribution function in the direction that corresponds to
the mid-point of the line connecting two neighboring vertices of an
icosahedron and a dodecahedron, defining thus $g_0(r)$. This correlation
function is included in \ref{SI_fig_gigd}{\bf a} as well and we see that it
shows significantly smaller oscillations than $g(r)$, a result that
is expected since one probes the structure in a direction which does
not pass close to the locations that correspond to the vertices of
the icosahedra/dodecahedra.  Panel {\bf b} shows that the length scale
over which $g(r)$, $g_I(r)$, and $g_D(r)$ decay is basically independent of the
function considered, demonstrating that they are indeed closely related
to each other.\\

\noindent
[30] S. Plimpton,
Fast parallel algorithms for short-range molecular dynamics.
{\it J. Comp. Phys.} {\bf 117}, 1 (1995).

\clearpage

\clearpage

\clearpage

\clearpage

\clearpage

\clearpage
\begin{figure}[ht]
\center
\includegraphics[width=0.45\columnwidth]{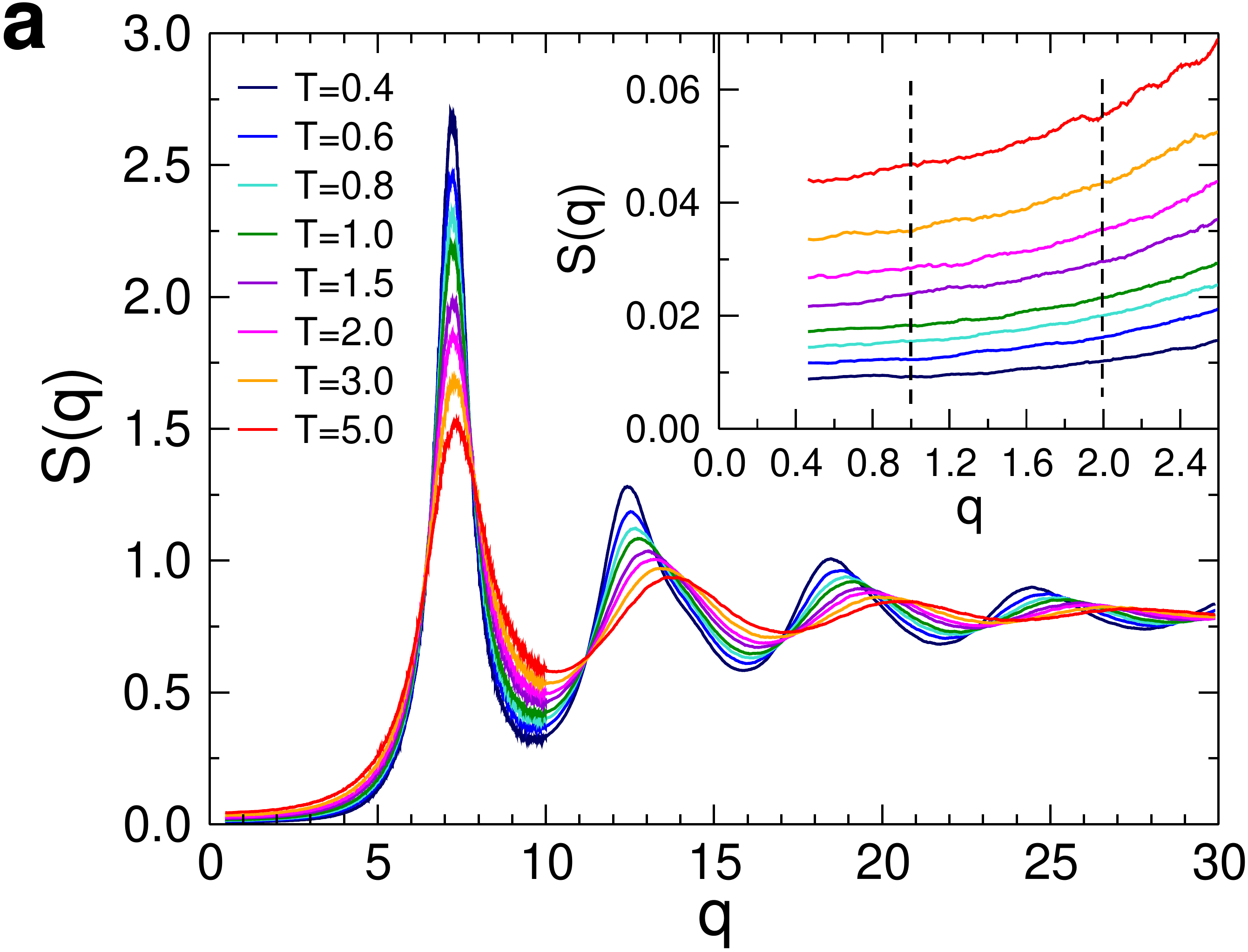}
\includegraphics[width=0.45\columnwidth]{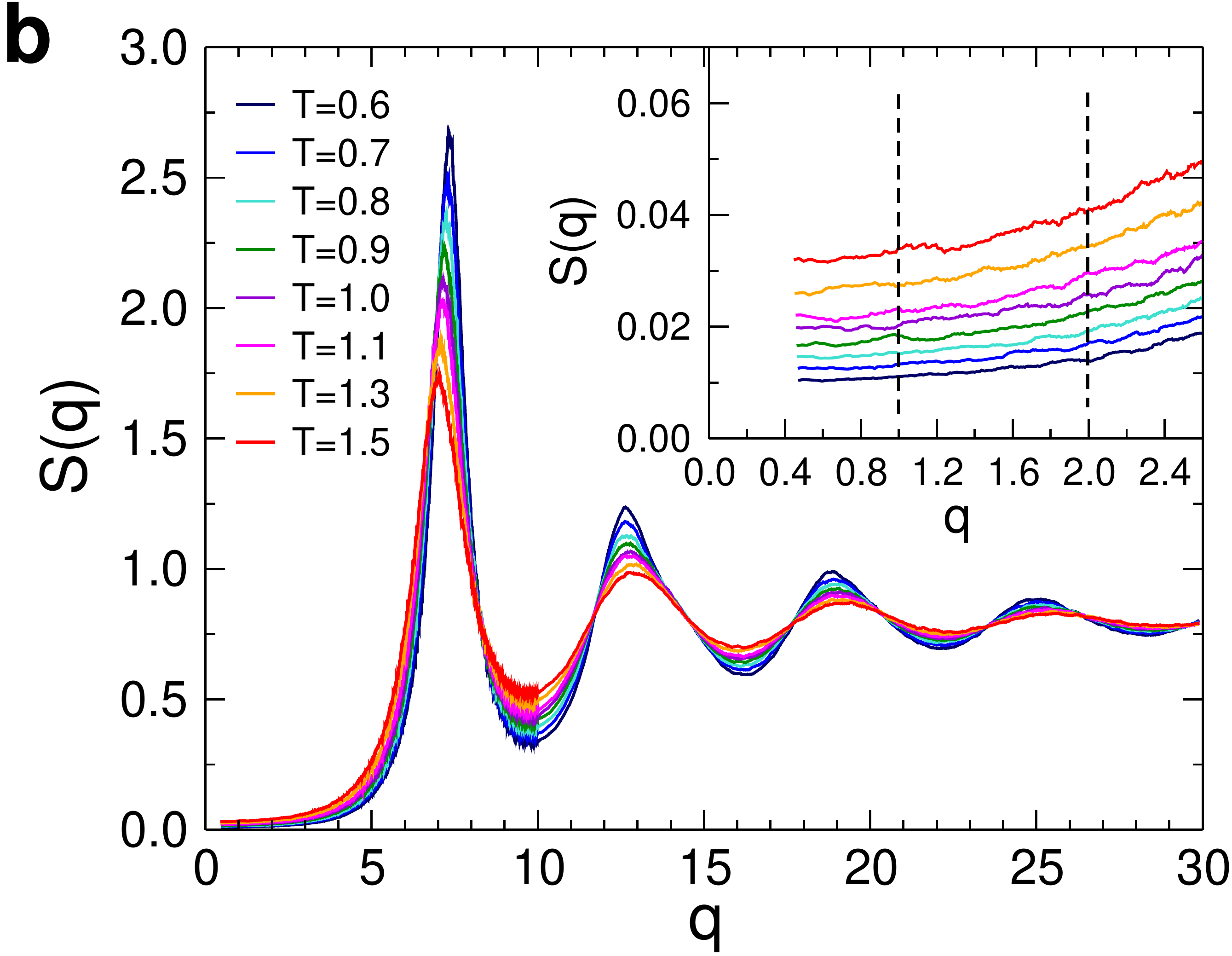}
\includegraphics[width=0.45\columnwidth]{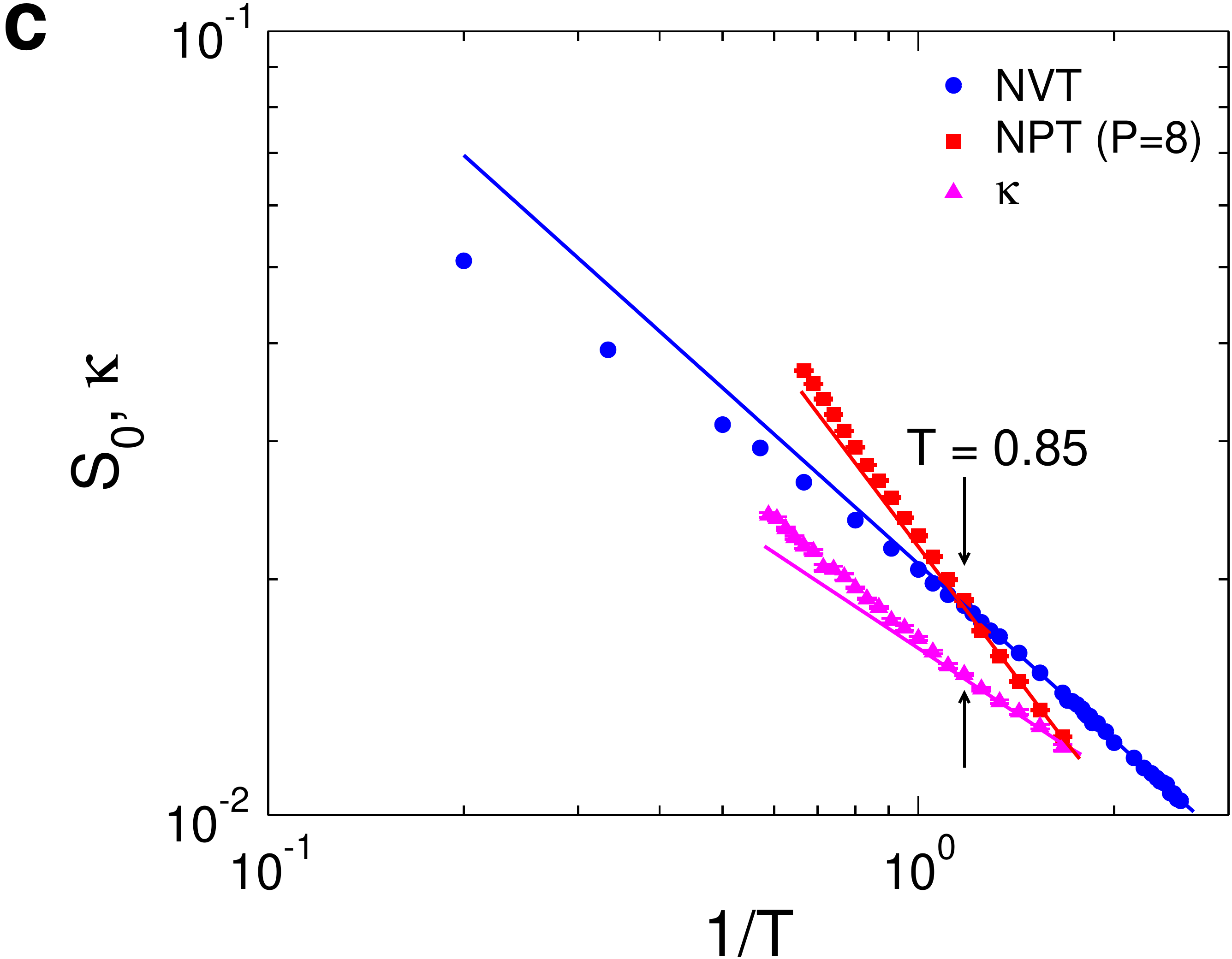}
\includegraphics[width=0.45\columnwidth]{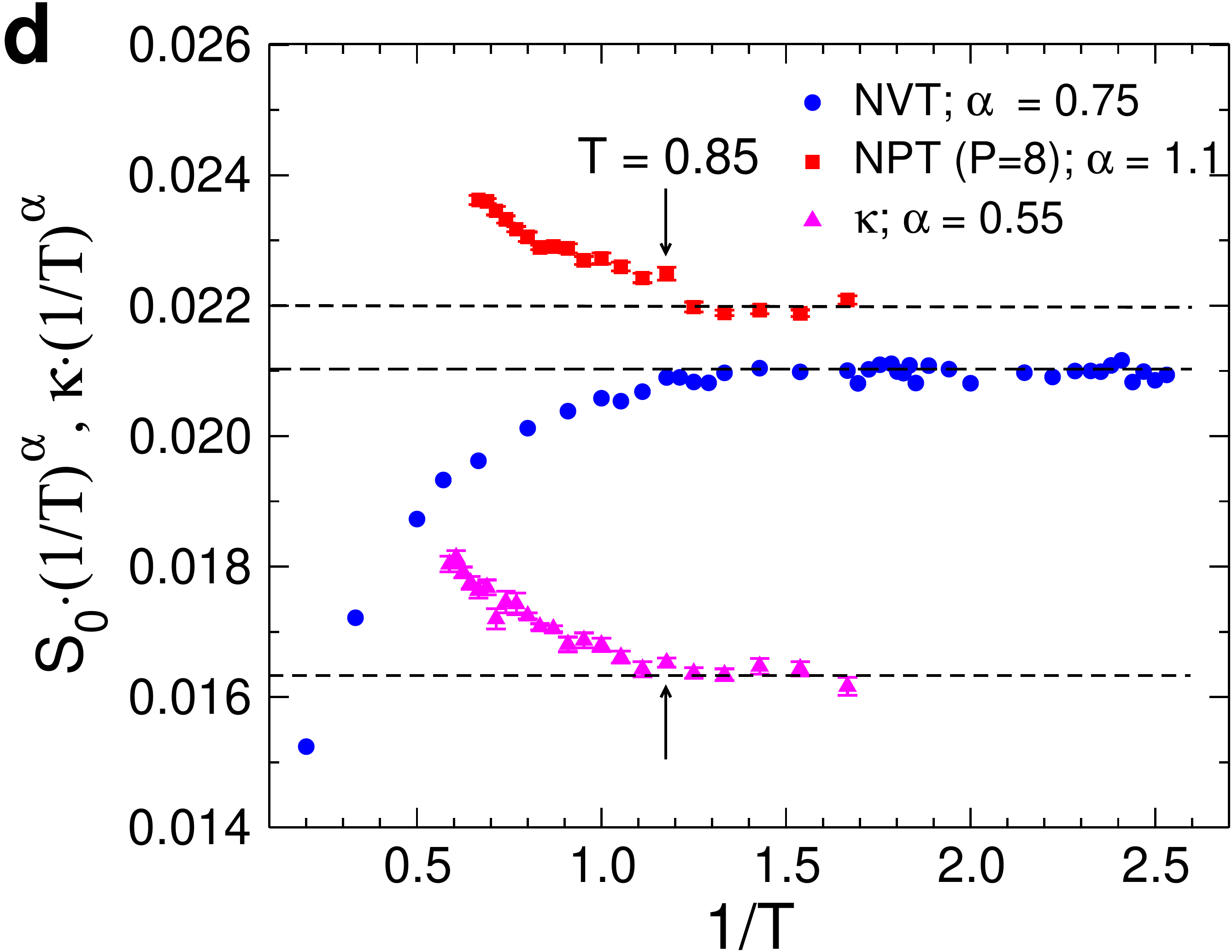}
\caption{{\bf Structure factor and compressibility.}
{\bf a} and {\bf b}: Partial static structure factor $S(q)$ for the AA pairs for simulations
at constant volume, {\bf a}, and constant pressure, {\bf b}. In panel {\bf b} the pressure is equal 
to 8.0. The insets show $S(q)$ at small $q$. The two vertical dashed
lines indicate the interval over which $S(q)$ was averaged in order to
obtain the value $S_0(T)$ representing $S(q)$ at small $q$. {\bf c}:
$S_0(T)$ as obtained for the two ensembles as a function of inverse
temperature. The magenta triangles are the compressibility from the NPT simulations. 
The solid lines are fits to the low-temperature data with
a power-law. {\bf d}: Same data as in panel {\bf c}, now multiplied by $(1/T)^\alpha$,
where the value of $\alpha$ is given in the legend. The horizontal dashed lines are
guides to the eye to see better that the three data sets show at around $T =
0.85$ a crossover  in their $T-$dependence.
}
\label{SI_fig_sq}
\end{figure}

\clearpage
\begin{figure}[ht]
\center
\includegraphics[width=0.7\columnwidth]{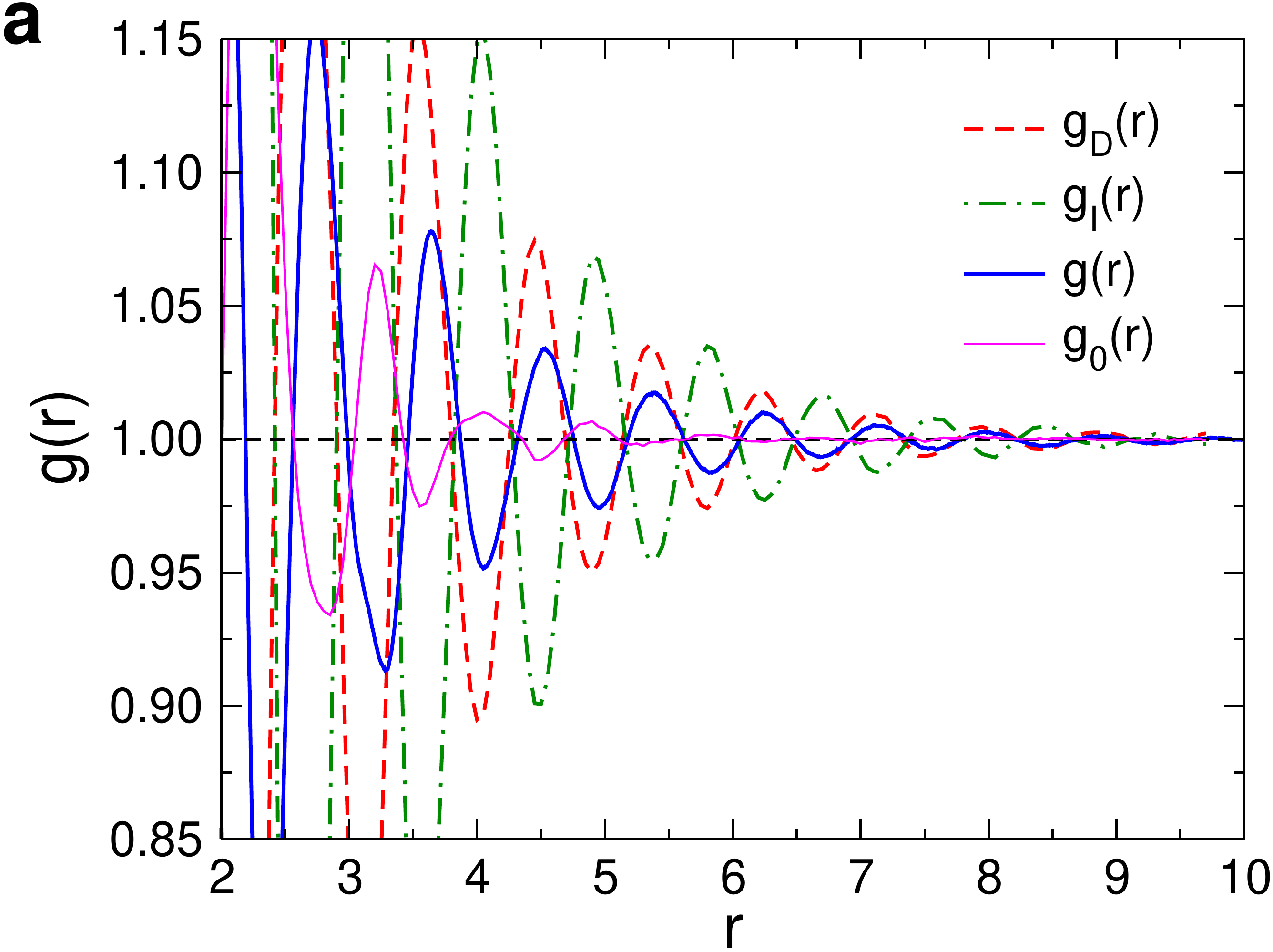}
\includegraphics[width=0.7\columnwidth]{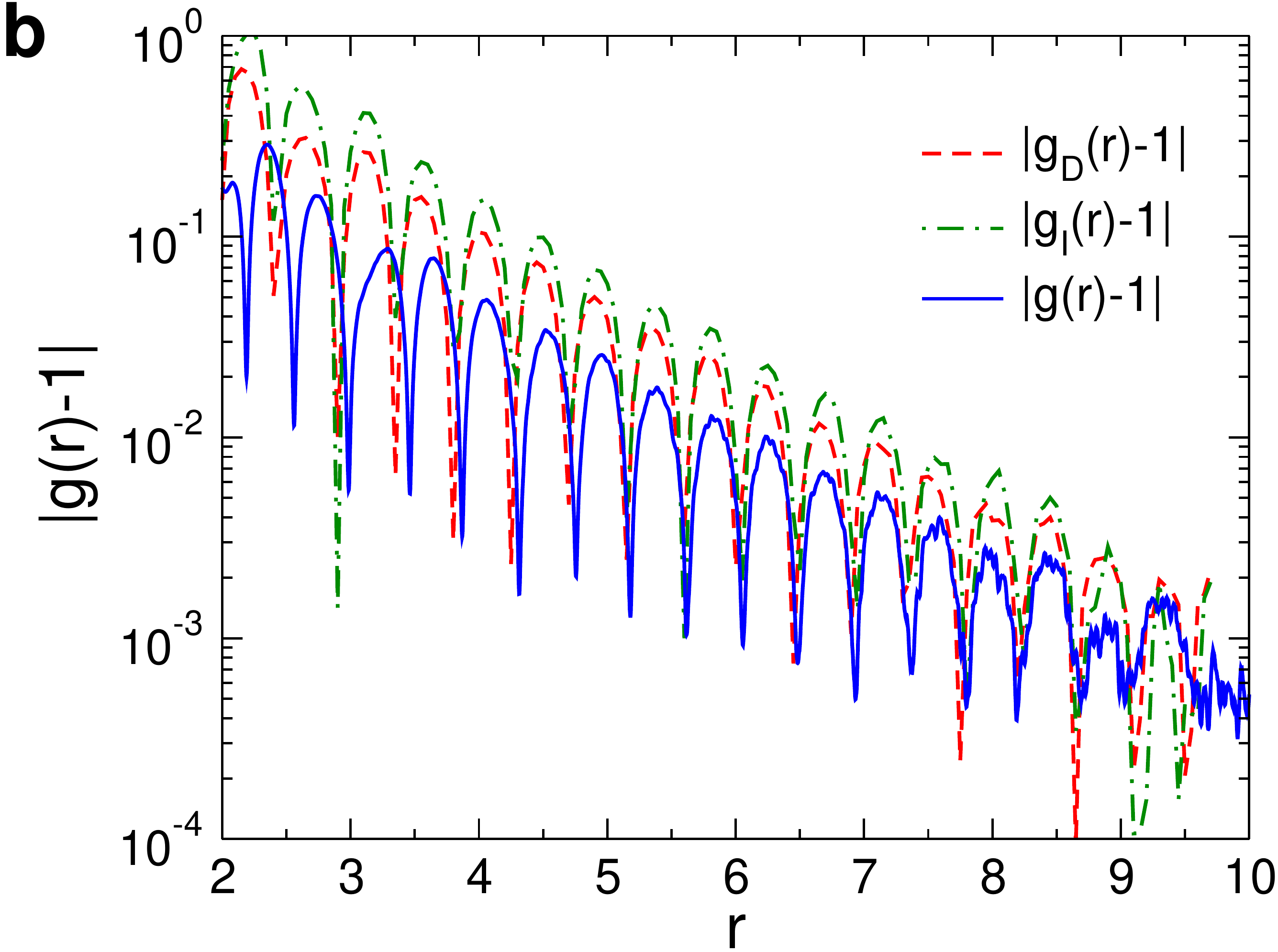}
\caption{{\bf Anisotropic radial distribution function.}
Radial distribution function as measured in the direction of the vertices
of the icosahedra, $g_I(r)$ and the direction of the vertices of the
dodecahedra, $g_D(r)$. The function $g_0(r)$ probes the structure in
the direction corresponding to the mid-point of the line connecting two neighboring vertices of an
icosahedron and a dodecahedron. Panels {\bf a} and {\bf b} show
these functions on a linear and log-scale, respectively. Also included
is $g(r)$, i.e., the radial distribution function averaged over all
directions.
}
\label{SI_fig_gigd}
\end{figure}

\clearpage
\noindent{ \bf Movie S1.}\\
This movie shows the density distribution $\rho(\theta,\phi,r)$ as a function
of the distance $r$ (left panel).  The right panel shows the partial
radial distribution function for A-N pairs (blue curve) as well as the
normalized angular power spectrum $S_\eta(6,r)$ (red curve). The center of the vertical
moving bar indicating the radius $r$ shown in the left panel. The
temperature is $T=2.0$.\\[5mm]

\noindent{ \bf Movie S2.}\\
This movie shows the density distribution $\rho(\theta,\phi,r)$ as a function
of the distance $r$ (left panel).  The right panel shows the partial
radial distribution function for A-N pairs (blue curve) as well as the
normalized angular power spectrum $S_\eta(6,r)$ (red curve). The center of the vertical
moving bar indicating the radius $r$ shown in the left panel. The
temperature is $T=0.4$.


\begin{thebibliography}{10}

\bibitem{binder_11}
K. Binder and W. Kob,
{\it Glassy Materials and Disordered Solids: An Introduction
to Their Statistical Mechanics}
(World Scientific, Singapore, 2005).

\bibitem{cosgrove_05}
D. J. Cosgrove,
Growth of the plant cell wall,
{\it Nat. Rev. Mol. Cell Biol. {\bf 6}, 850 (2005)}.

\bibitem{becker_13}
S. M. Becker and A. V. Kuznetsov,
{\it Transport in Biological Media}
(Academic Press, Amsterdam, 2013).

\bibitem{ashcroft_76}
N. W. Ashcroft and N. D. Mermin, 
{\it Solid State Physics} 
(Holt-Saunders, New York, 1976).

\bibitem{hansen_86}
J. P. Hansen and I. R. McDonald, 
{\it Theory of Simple Liquids}
(Elsevier, Amsterdam, 1986).

\bibitem{salmon_06}
P. S. Salmon, 
Decay of the pair correlations and small-angle scattering for binary liquids and glasses.
{\it J. Phys.: Condens. Matter.} {\bf 18}, 11443 (2006).

\bibitem{tanaka_12}
M. Leocmach and H. Tanaka, 
Roles of icosahedral and crystal-like order in the hard spheres glass transition.
{\it Nat. Comm.} {\bf 3}, 974 (2012).

\bibitem{royall_15}
C. P. Royall and S. R. Williams,
The role of local structure in dynamical arrest.
{\it Phys. Rep.} {\bf 560}, 1 (2015).

\bibitem{schoenholz_16}
S.S. Schoenholz, E.D. Cubuk, D.M. Sussman, E. Kaxiras, and A.J. Liu,
A structural approach to relaxation in glassy liquids.
{\it Nat. Phys.} {\bf 12}, 469 (2016).

\bibitem{royall_17}
C. P. Royall and W. Kob, 
Locally favoured structures and dynamic length scales in a simple glass-former.
{\it J. Stat. Mech.: Theo. Exp.} {\bf 2}, 024001 (2017).

\bibitem{coslovich_07}
D. Coslovich and G. Pastore, 
Understanding fragility in supercooled Lennard-Jones mixtures. I. Locally preferred structures.
{\it J. Chem. Phys.} {\bf 127}, 124504 (2007).

\bibitem{dunleavy_15}
A. J. Dunleavy, K. Wiesner, R. Yamamoto, and C. P. Royall,
Mutual information reveals multiple structural relaxation mechanisms in a model glass former.
{\it Nat. Comm.} {\bf 6}, 6089 (2015).

\bibitem{jonsson_88}
H. Jonsson and H. C. Andersen, 
Icosahedral ordering in the Lennard-Jones liquid and glass.
{\it Phys. Rev. Lett.} {\bf 60}, 2295 (1988).

\bibitem{wochner_09}
P. Wochner, C. Gutt, T. Autenrieth, T. Demmer, V. Bugaev, A. D. Ortiz, A. Duri, F. Zontone, 
G. Gr\"ubel, and H. Dosch,
{\it Proc. Natl. Acad. Sci. USA} {\bf 106}, 11511 (2009).

\bibitem{malins_13}
A. Malins, J. Eggers, C. P. Royall, S. R. Williams, and H. Tanaka, 
Identification of long-lived clusters and their link to slow dynamics in a model glass former.
{\it J. Chem. Phys.} {\bf 138}, 12A535 (2013).

\bibitem{ma_11}
Y. Q. Cheng and E. Ma,
Atomic-level structure and structure-property relationship in metallic glasses.
{\it Prog. Mater. Sci.} {\bf 56}, 379 (2011).

\bibitem{miracle_04}
D. B. Miracle, A structural model for metallic glasses.
{\it Nat. Mater.} {\bf 3}, 697 (2004).

\bibitem{xia_17}
C. Xia, J. Li, B. Kou, Y. Cao, Z. Li, X. Xiao, Y. Fu, T. Xiao, L. Hong, J. Zhang, W. Kob, and Y. Wang,
Origin of non-cubic scaling law in disordered granular packing.
{\it Phys. Rev. Lett.} {\bf 118},  238002 (2017).
 
\bibitem{adam_65}
G. Adam and J. H. Gibbs, 
On the temperature dependence of cooperative relaxation properties in glass-forming liquids.
{\it J. Chem. Phys.} {\bf 43}, 139 (1965).

\bibitem{rfot}
X. Y. Xia and P. G. Wolynes, 
Fragilities of liquids predicted from the random first order transition theory of glasses.
{\it Proc. Natl. Acad. Sci. USA}. {\bf 97}, 2990 (2000).

\bibitem{chandler_10}
D.~Chandler and J.~P. Garrahan,
Dynamics on the way to forming glass: Bubbles in space-time.
{\it Annual Review of Physical Chemistry}, 61, 191 (2010).

\bibitem{fang_10}
X. W. Fang, C. Z. Wang, Y. X. Yao, Z. J. Ding, and K. M. Ho,
Atomistic cluster alignment method for local order mining in liquids and glasses.
{\it Phys. Rev. B} {\bf 82}, 184204 (2010).

\bibitem{fang_11}
X. W. Fang, C. Z. Wang, S. G. Hao, M. J. Kramer, Y. X. Yao,
M. I. Mendelev, Z. J. Ding, R. E. Napolitano, and K. M. Ho,
Spatially resolved distribution function and the medium-range order in metallic liquid and glass.
{\it Scient. Reps.} {\bf 1}, 194 (2011).

\bibitem{kob_95}
W.~Kob and H.~C. Andersen, 
Testing mode-coupling theory for a supercooled binary Lennard-Jones mixture I: The van Hove correlation function.
{\it Phys. Rev. E} {\bf 51}, 4626 (1995).

\bibitem{mct}
W. G\"otze, {\it Complex dynamics of glass-forming liquids:
A mode-coupling theory} (Oxford University Press, Oxford, 2008).

\bibitem{kegel_00}
W. K. Kegel and A. van Blaaderen,
Direct Observation of Dynamical Heterogeneities in Colloidal Hard-Sphere Suspensions.
{\it Science} {\bf 287}, 290 (2000).

\bibitem{weeks_00}
E. R. Weeks, J. C. Crocker, A. C. Levitt, A. Schofield, and D. A. Weitz,
Three-dimensional direct imaging of structural relaxation near the colloidal glass transition.
{\it Science} {\bf 287}, 627 (2000).

\bibitem{sherson_10}
J. F. Sherson, C. Weitenberg, M. Endres, M. Cheneau1, I. Bloch, and S. Kuhr,
Single-atom-resolved fluorescence imaging of an atomic Mott insulator.
{\it Nature} {\bf 467}, 68 (2010).

\bibitem{kou_17}
B. Kou, Y. Cao, J. Li, C. Xia, Z. Li, H. Dong, A. Zhang, J. Zhang, W. Kob, and Y. Wang
Granular materials flow like complex fluids.
{\it Nature} {\bf 551}, 360 (2017).


\end{thebibliography}
\end{document}